\documentclass[prl,aps,reprint]{revtex4-1}

\usepackage{graphicx}
\usepackage[english]{babel}
\usepackage{amsmath}
\usepackage{amsfonts}
\usepackage{amssymb}
\usepackage[squaren]{SIunits}
\usepackage{rotfloat}

\begin{document}

\title{Inter-valley dark trion states with spin lifetimes of 150 ns in WSe$_2$}

\author{F. Volmer}
\author{S. Pissinger}
\author{M. Ersfeld}
\author{S. Kuhlen}
\author{C. Stampfer}
\author{B. Beschoten}
\thanks{E-mail address: bernd.beschoten@physik.rwth-aachen.de}
\affiliation{2nd Institute of Physics and JARA-FIT, RWTH Aachen University, D-52074 Aachen, Germany}

\date{\today}

\begin{abstract}

We demonstrate long trion spin lifetimes in a WSe$_2$ monolayer of up to \unit{150}{ns} at \unit{5}{K}. Applying a transverse magnetic field in time-resolved Kerr-rotation measurements reveals a complex composition of the spin signal of up to four distinct components. The Kerr rotation signal can be well described by a model which includes inhomogeneous spin dephasing and by setting the trion spin lifetimes to the measured excitonic recombination times extracted from time-resolved reflectivity measurements. We observe a continuous shift of the Kerr resonance with the probe energy, which can be explained by an adsorbate-induced, inhomogeneous potential landscape of the WSe$_2$ flake. A further indication of extrinsic effects on the spin dynamics is given by a change of both the trion spin lifetime and the distribution of $g$-factors over time. Finally, we detect a Kerr rotation signal from the trion's higher-energy triplet state when the lower-energy singlet state is optically pumped by circularly polarized light. We explain this by the formation of dark trion states, which are also responsible for the observed long trion spin lifetimes.

\end{abstract}

\maketitle

As direct band gap semiconductors, monolayers of transition metal dichalcogenides (TMDCs) allow an interesting and promising extension for the family of two-dimensional (2D) materials. They fill the gap between the zero band gap, high mobility graphene and the insulating hexagonal boron nitride, which paves the way for advanced devices with tailored physical properties by means of so-called van-der-Waals heterostructures \cite{Nature.499.419}. The unique feature of TMDCs within the class of 2D materials consists of their large valley-dependent spin-orbit splitting combined with optical selection rules, which enables valley-selective optical excitation of electron and hole spins \cite{PhysRevLett.108.196802, NatPhys.10.343}. So far several studies exploited this unique feature to investigate the TMDCs' spin dynamics by optical pump-probe experiments of the time-resolved Faraday rotation (TRFR) or Kerr rotation (TRKR) \cite{NatureComm.6.896,PhysRevB.90.161302,NanoLett.16.5010,NatureComm.7.12715,NatPhys.11.830,NanoLett.15.8250,PhysRevB.92.235425,arXiv150704599Y,arXiv160203568B,arXiv161201336Y}. Interestingly, these studies have yielded quite opposing results: 1) The measured time constants of the Kerr or Faraday signals differ by more than three orders of magnitude from some ps to tens of ns. 2) The signals can sometimes be described by a single exponential decay, whereas others show multiple components. 3) The magnetic field dependence can either show clear oscillatory behavior by Larmor spin precession, an inhomogeneous dephasing of the optically-induced spin ensemble, or is non-existent at all. 4) And finally, the spin lifetimes measured in Kerr or Faraday signal is reported to be either shorter, equal, or even longer than the exciton recombination time, which is determined either from time-resolved photoluminescence (TRPL) or from time-resolved reflectivity (TRR) measurements. In particular the fact about different lifetimes measured in spin-sensitive TRKR/TRFR and charge-sensitive TRPL/TRR measurements led to quite different conclusions about the origin of the detected spin signals, as some studies conclude a transfer of spin polarization to resident carriers in the TMDC, whereas other studies see no evidence for this transfer (an overview of the different studies and their respective results is provided in the supplementary information~\cite{Supplement}). Here, we present extrinsic effects as an explanation for these quite opposing results. The entirety of our observations suggests the possibility that the spin properties of TMDCs' monolayers can be significantly and systematically tailored by modifying and controlling their environment, in our case e.g. by adsorbates.

\begin{figure*}[tb]
	\includegraphics{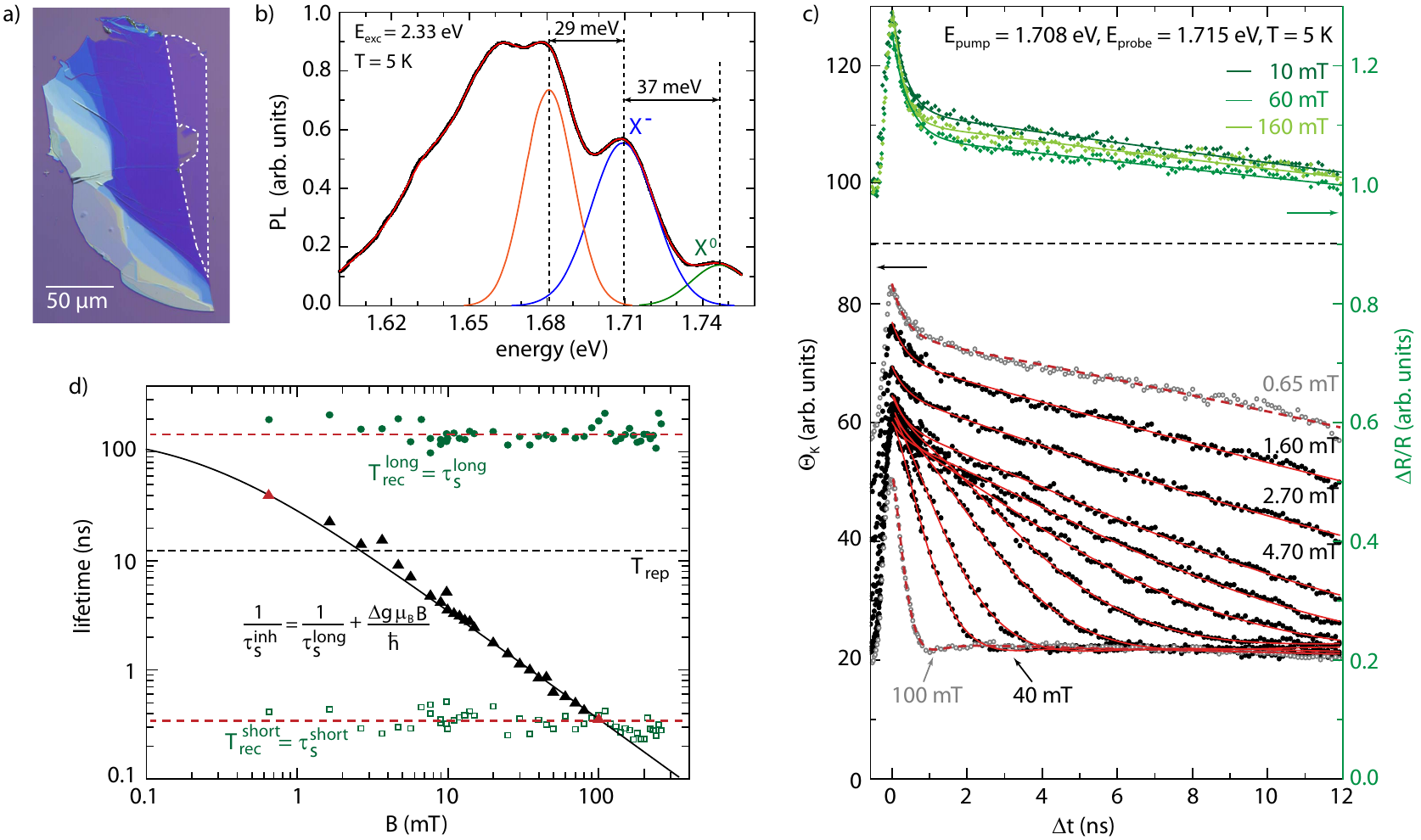}
	\caption{(a) Optical image of the exfoliated WSe$_2$ flake on SiO$_2$. The monolayer part is marked by the white dashed line. (b) Photoluminescence spectrum on the monolayer part at \unit{5}{K} and an excitation wavelength of \unit{532}{nm}. The black line represents the data and the red one the sum of the fitted Gaussian functions. The energy separation between the neutral exciton (X$^0$) and trion peak (X$^-$) is quite large ($\unit{37\pm1}{meV}$) which is an indication for a highly n-doped sample \cite{NatNano.8.634, NatNano.9.268}. The peaks at lower energies which exhibit the highest photoluminescence (PL)intensities are assigned to localized excitonic states. (c) Time-resolved reflectivity (upper part, green curves) and Kerr-rotation $\Theta_\text{K}$ (lower part, grayish-black curves). All curves are plotted as measured and hence the large signal at negative delays ($\Delta t<0$, shortly before the arrival of the pump pulse) shows very long-lived recombination and trion spin states which lifetimes significantly exceed the laser repetition interval of $T_{\text{rep}}=\unit{12.5}{ns}$. The reflectivity signal exhibits a short ($T_{\text{rec}}^{\text{short}}$) and a long exponential decay time ($T_{\text{rec}}^{\text{long}}$) with no magnetic field dependence (d). The Kerr-signal, on the other hand, shows three components: A long-lived ($>T_{\text{rep}}$) and short-lived ($<\unit{1}{ns}$) decay time. Both are magnetic field independent and can be fitted by assuming that their lifetimes are equal to the two decay times seen in the reflectivity measurements. The third component depends inversely on the magnetic field which results from an inhomogeneous spin dephasing mechanism.}
	\label{fig1}
\end{figure*}

We report on time- and energy-resolved Kerr-rotation measurements on a monolayer WSe$_2$ flake. The investigated sample shows quite complex spin dynamics of four different spin components with varying lifetimes up to \unit{150}{ns} when resonantly exciting into charged exciton (trion) states far below the fundamental band gap of WSe$_2$. In a magnetic field, which is applied transverse to the spin orientation, two components show no magnetic field dependence at all, whereas one spin component undergoes inhomogeneous dephasing and the last one shows a clear oscillatory behavior resulting from Larmor spin precession. The different lifetimes in TRKR measurements are also seen in TRR measurements. The latter can be utilized to monitor the exciton recombination process. The limitation of the measured spin lifetime in TRKR to the excitonic recombination indicates that the long spin lifetimes do not result from the spin polarization of resident charge carriers but are rather due and restricted to the spin polarization of the excitonic states. To avoid confusion with the one-particle spin lifetimes of free charge carriers, we therefore denote the extracted lifetimes from the Kerr measurements as excitonic spin lifetimes. To gain further insight into the excitonic spin dynamics, we employ a system of two synchronized pulsed lasers which allows us to independently tune the energies of both pump and probe pulses through the fine-structure of the trion which includes its singlet and triplet state. The corresponding Kerr rotation amplitude exhibits an energy splitting of this fine-structure of \unit{7}{meV} which is consistent to theoretical predictions. Interestingly, we measure a Kerr signal from the trion's higher-energy triplet state when the lower-energy singlet state is pumped. To explain this observation, we introduce a model in which the formation of higher-energy, bright exciton states is blocked by dark states, which are populated through inter-valley relaxation processes of the pumped excitonic states. These negatively charged dark trion states are most likely also the reason for the observed long excitonic spin lifetimes. We argue that the formation of these dark trion states require an efficient momentum scattering mechanism, which is provided in our sample by a high concentration of adsorbates and localized states. Another fingerprint that extrinsic effects dominate the spin dynamics in our sample is the observation that the Kerr resonance shifts continuously with the probe energy, which we explain by an adsorbate-induced, inhomogeneous potential landscape along the WSe$_2$ flake.

For both TRKR \cite{PhysRevB.56.7574} and TRR experiments we use two mode-locked Ti:sapphire lasers to independently tune the energies of both pump and probe pulses. An electronic delay between both pulses covers the full laser repetition interval of \unit{12.5}{ns} with a jitter of less than \unit{1}{ps}. The pulse width is \unit{3}{ps} for the pump pulses and \unit{100}{ps} for the probe pulses. Both pulse trains are focused to a spot size of approximately $\unit{30}{\mu m}$ and their power was kept in a range of $\unit{500-800}{\mu W}$ each. A detailed scheme of the experimental setup can be found in Ref.~\cite{Supplement}. The investigated WSe$_2$ flake was mechanically exfoliated with a polymer-stamp (PDMS) from a bulk crystal and transferred onto a Si/SiO$_2$-substrate. The monolayer part of the flake, on which all measurements were conducted, is marked in the optical image of Fig.~\ref{fig1}(a) by a dashed line and was confirmed by Raman spectroscopy \cite{Supplement}. The monolayer was also characterized by photoluminescence (PL) spectroscopy at \unit{5}{K} and an excitation energy of \unit{2.33}{eV} (Fig.~\ref{fig1}(b)). The PL spectrum was fitted by Gaussian peak functions (only the 3 high energy peaks are included in Fig.~\ref{fig1}(b)) and exhibits a quite large energy separation of $\unit{37\pm1}{meV}$ between the neutral exciton (X$^0$, $E\approx\unit{1.746}{eV}$) and charged exciton (trion) peak (X$^-$, $E\approx\unit{1.709}{eV}$). Usually, a trion binding energy of $\unit{30}{meV}$ is reported in literature \cite{NatPhys.12.323, PhysRevB.90.161302, ScientificRep.6.22414, PhysRevB.90.075413}, but this value is only valid at low charge carrier concentrations as a charge carrier dependent red shift of the trion peak is observed in case of WSe$_2$ \cite{NatNano.8.634, NatNano.9.268}. According to Ref.~\cite{NatNano.8.634} the large binding energy is an indication for a highly n-doped sample, because the positively charged trion does not show such a high redshift with charge carrier density. The additional emission peaks in the lower-energy branch of the PL spectrum are attributed to bound excitons at localized states \cite{NatPhys.11.477, APL.105.101901, ScientificRep.6.22414, PhysRevX.6.021024, PhysRevB.90.075413}. The PL-intensities of these peaks surpass the one of the trion and especially the one of the neutral exciton peak. This indicates a significant concentration of localized states which provide momentum scattering centers. Later on we will argue that these scattering centers are a necessary requirement for the formation of dark states and therefore the long excitonic spin lifetimes.

One important aspect for the understanding of the spin dynamics and the long spin lifetimes in the WSe$_2$ flake is revealed from time-resolved reflectivity measurements. This measurement technique is used to probe the recombination dynamics in semiconductors as the decay times of the reflectivity signal can be assigned to the lifetimes of excitonic recombinations \cite{PhysRevB.90.161302, NatureComm.6.896, NatureComm.7.12715, NatPhys.11.830, PhysRevB.92.235425, arXiv150704599Y}. The reflectivity curves in the upper part of Fig.~\ref{fig1}(c) (green curves) are measured in the trion regime at $T=\unit{5}{K}$ and plotted as measured (see corresponding scale bar on the right axis). It is apparent that the TRR signal only weakly decays on the $\unit{12.5}{ns}$ repetition interval of the laser system. This results in large values of the pump beam-induced reflectivity change $\Delta R/R$ at negative delay times ($\Delta t<0$, i.e. shortly before the arrival of the next pump pulse at $\Delta t=0$) indicating a very long-lived exciton recombination time on the order of tens of nanoseconds. This observation is in contrast to many reported values in literature about exciton dynamics as, e.g., TRPL measurements usually exhibit decay times in the ps range both for WSe$_2$ \cite{PhysRevB.90.075413, APL.105.101901,NatPhys.11.477, PhysRevB.93.205423} and also for other TMDCs like MoSe$_2$, WS$_2$, or MoS$_2$ \cite{PhysRevLett.112.047401, PhysRevB.93.205423,Nanoscale.7.7402}. Now it is important to note that PL measurements only detect the recombination of bright excitons (i.e. electron-hole recombinations which are followed by the emission of photons). On the other hand, the TRR measurements (and later on also the TRKR measurements) are sensitive to the optically-induced imbalance in the population of electron and hole states by the pump pulse. But such an imbalance can also be caused by so-called dark neutral or charged exciton states that do not recombine by the emission of a photon and are therefore not visible in PL measurements \cite{NatSciRev.2.57}. There already exist direct experimental evidence for such dark states \cite{Nature.513.214, PhysRevLett.115.257403} and several features in optical measurements on TMDCs were assigned to these states \cite{PhysRevB.93.205423, NatMater.14.889, Nanoscale.7.10421}. We attribute the extraordinary long exciton recombination times extracted from the TRR measurements to the formation of dark trion states, which explains many features in the TRKR measurements later on in this study. Next to the extraordinary long recombination lifetimes, we also observe an additional recombination on the time-scale of some hundred ps. The occurrence of multiple decay times is already observed in other reflectivity measurements \cite{ACSNano.8.2970,NatureComm.6.896,NatureComm.7.12715,PhysRevB.92.235425} and the short-lived signal may represent the time-scale over which optically excited bright trion states either scatter into the dark trion states or radiatively recombine.

When fitting the time constants from the TRR data, we took into account that the long-lived recombination time is much longer than the laser repetition interval $T_{\text{rep}}$. Therefore, the influence of the subsequent pump pulse on an already existing exciton population stemming from the previous pulses has to be considered. As we discus in the supplement \cite{Supplement}, we assume that the pump pulse will not disturb the already existing excitons by laser beam-induced effects such as band gap renormalization \cite{NatPhoton.9.466} or the Stark effect \cite{Science.346.1205} as the pulse fluence in our measurements is quite low (around $\unit{1}{\mu Jcm^{-2}}$). Accordingly, the TRR data can be fitted by a sum over several pulses and a bi-exponential decay of the form:
\begin{eqnarray}
	\frac{\Delta R}{R}&= &A_{\text{rec}}^{\text{short}} \text{exp} \left( -\frac{t}{T_{\text{rec}}^{\text{short}}} \right) \nonumber \\
	&+& \sum\limits_n  A_{\text{rec}}^{\text{long}} \text{exp} \left( -\frac{t+n  T_{\text{rep}}}{T_{\text{rec}}^{\text{long}}} \right), \quad t>0
	\label{eq:rec}
\end{eqnarray}
with each a long and short recombination time $T_{\text{rec}}^{\text{i}}$ with their respective amplitude $A_{\text{rec}}^{\text{i}}$. The analysis of the TRR data by Eq.~\ref{eq:rec} yields long trion recombination times of around $\unit{150}{ns}$. We measured the reflectivity curves for different magnetic fields (see selected curves in Fig.~\ref{fig1}(c)) applied in the plane of the WSe$_2$ flake but perpendicular to the laser beams and observe no discernible field dependence for both the long ($\unit{150}{ns}$) and short ($\unit{380}{ps}$) recombination times, which is seen in Fig.~\ref{fig1}(d). Slight variations between different magnetic field values (as plotted in the upper panel of Fig.~\ref{fig1}(c)) can be explained by the fact that the whole cryostat and therefore also the spot position on the flake slightly moves when changing the magnetic field. This is important as PL and TRKR measurements already demonstrated that a TMDC flake can have spatially-varying optical properties \cite{arXiv160203568B}. As we explain further below, we attribute these variations to an inhomogeneous distribution of adsorbates on top of the TMDC flake.

Next, we discuss the TRKR measurements which are shown in the lower panel of Fig.~\ref{fig1}(c). These traces were recorded under identical experimental conditions as the TRR data. We first focus on the curve for the lowest magnetic field (\unit{0.65}{mT}). Both at short and long delay times this curve exhibits a similar temporal evolution as the TRR curves. This suggests that the spin dynamics seen in TRKR is closely linked to the exciton dynamics probed by TRR. Interestingly, there is a strong magnetic field dependence in the TRKR data, which shows a decreasing spin lifetime with increasing magnetic field as seen by the increasing slope of $\Theta_K$ which merges into the temporal regime of several hundred ps of the fast exciton dynamics at larger magnetic fields ($\unit{100}{mT}$). Nevertheless, even at these larger magnetic fields there is a pronounced Kerr rotation signal at long delay times which does not depend on the magnetic field and which shows only a slight decay over the measured timescale.

The measured Kerr rotation $\Theta_\text{K}$ can be fitted under the assumption of a magnetic field independent short- and long-lived spin component ($\tau_{\text{s}}^{\text{short}}$ and $\tau_{\text{s}}^{\text{long}}$, respectively) and a third spin component ($\tau_{\text{s}}^{\text{inh}}$), which undergoes Larmor spin precession:
\begin{eqnarray}
\Theta_{\text{K}}&=& A_{\text{s}}^{\text{short}} \text{exp} \left( -\frac{t}{\tau_{\text{s}}^{\text{short}}} \right)+ A_{\text{s}}^{\text{long}} \text{exp} \left( -\frac{t}{\tau_{\text{s}}^{\text{long}}} \right) \nonumber \\
&+& \Bigg[ \sum\limits_n  A_{\text{s}}^{\text{inh}} \text{cos} \left( \frac{g \mu_{\text{B}} B}{\hbar} (t+n  T_{\text{rep}})  + \phi \right) \nonumber \\
&& \text{exp} \left( -\frac{t+n  T_{\text{rep}}}{\tau_{\text{s}}^{\text{inh}}} \right) \Bigg], \quad t>0
\label{eq:threeExp}
\end{eqnarray}
with the Land\'e $g$-factor, the Bohr magneton $\mu_{\text{B}}$, a phase $\phi$ of the spin precession, and the respective spin amplitudes $A_{\text{s}}^{\text{i}}$. We note that a pure exponential term cannot describe the magnetic field dependent component alone. As we demonstrate in Ref.~\cite{Supplement}, the cosine term is needed to describe the curves in every detail. In Eq.~\ref{eq:threeExp} we also assume that successive pump pulses do not disturb the decay dynamics of the spin components (see discussion in Ref.~\cite{Supplement}). A very good fit to the data (red solid lines in Fig.~\ref{fig1}(c)) is achieved when we assume that the magnetic field independent short- and long-lived spin components have identical lifetimes to the trion recombination times determined from TRR, i.e. $T_{\text{rec}}^{\text{short}} = \tau_{\text{s}}^{\text{short}} = \unit{380}{ps}$ and $T_{\text{rec}}^{\text{long}} = \tau_{\text{s}}^{\text{long}} = \unit{150}{ns}$. For both we chose the averaged values over all magnetic field scans (dashed lines in Fig.~\ref{fig1}(d)).

\begin{figure*}[tb]
	\includegraphics{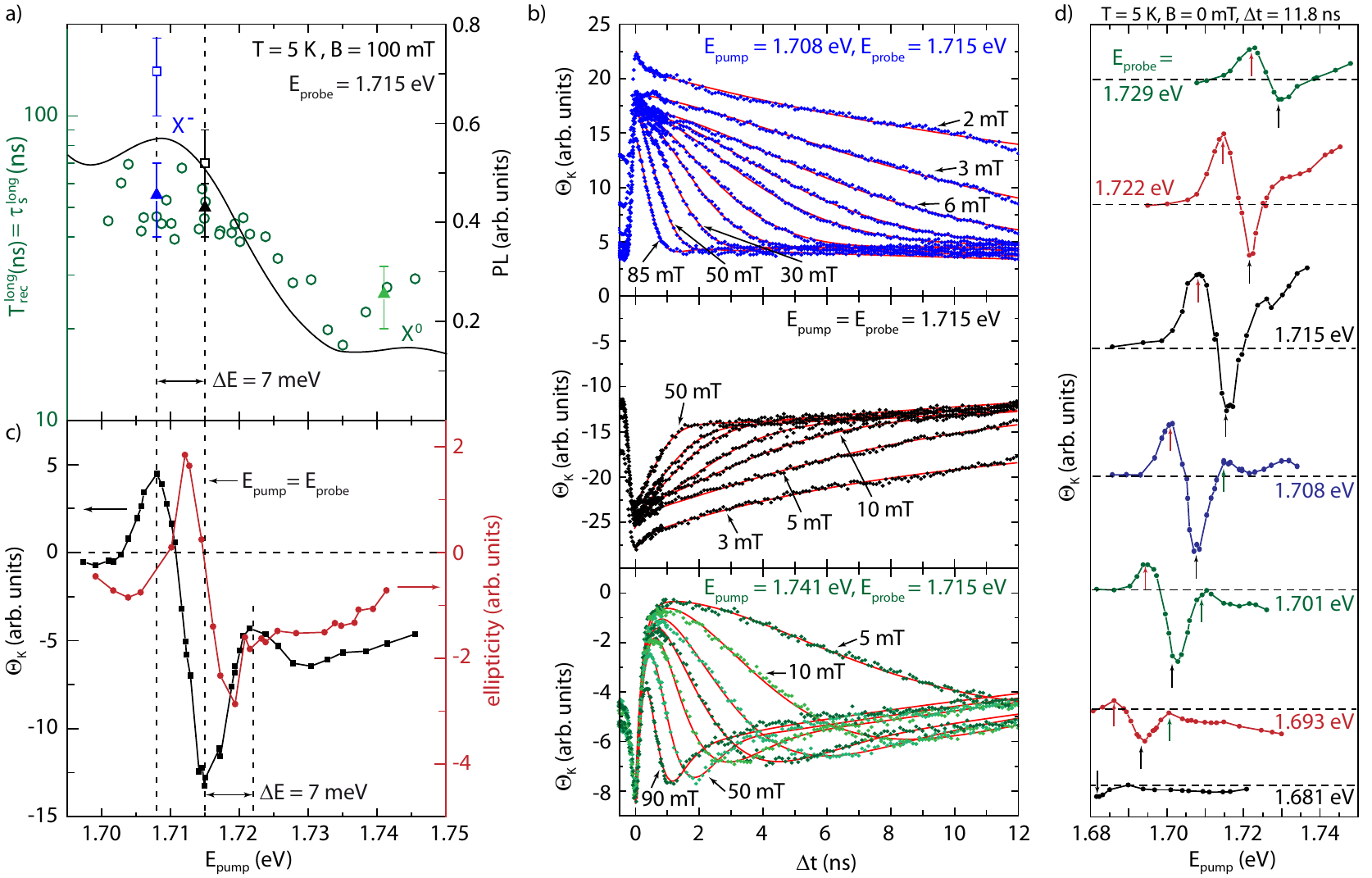}
	\caption{(a) The long-lived component of the exciton recombination as a function of pump energy between exciton and trion energy. A close-up of the PL spectrum from Fig.~\ref{fig1}(b) is included. The probe energy was set to $E_{\text{probe}}=\unit{1.715}{eV}$. (b) A complete set of magnetic field dependent Kerr-measurements were conducted for three different pump energies. The extracted lifetimes of the long-lived Kerr-component are given by the triangular symbols in (a) and match the decay times of the long-lived recombination signal. In case of two energies, we remeasured the magnetic field dependence in a later cooling-cycle and observed an increase in the measured lifetimes (quadratic symbols, the blue one represents the data-set of Fig.~\ref{fig1}(c)). The red curves in (b) are fits done with the inhomogeneous dephasing model explained in the main text (Eq.~\ref{eq:inh}). (c) Amplitudes of the long-lived Kerr- and ellipticity-signals of the pump-scan shown in (a). For $E_{\text{pump}}=E_{\text{probe}}$ the Kerr-rotation is maximal whereas the ellipticity-signal exhibits a zero-crossing. Interestingly, another maximum of the Kerr-signal can be measured if the pump energy is \unit{7}{meV} below the probe energy. (d) Changing the probe energy leads to a continuous shift of the whole resonance: There is always a local maximum for $E_{\text{pump}}=E_{\text{probe}}$ (indicated by black arrows, dashed lines represents $\Theta_K=0$) and another local maximum with inverted sign at a pump energy approximately \unit{7}{meV} below the probe energy (indicated by red arrows), which can be interpreted as the energy splitting between the trion's singlet and triplet state \cite{NatureComm.5.3876, NatureComm.7.12715}.}
	\label{fig2}
\end{figure*}

The magnetic field dependent spin lifetimes of the third component ($\tau_{\text{s}}^{\text{inh}}(B)$) which shows spin precession is also depicted in Fig.~\ref{fig1}(d) (triangles). Within this double-logarithmic plot the fitted spin lifetimes follow a $1/B$-dependence with a slope of $-1$. Such a magnetic field dependence has already been observed in TMDCs \cite{NanoLett.15.8250} and in other semiconductors \cite{PhysRevB.66.125307} and is attributed to an inhomogeneous spin dephasing mechanism, where a spin ensemble dephases in a transverse magnetic field as individual spins are subject to slightly different $g$-factors within a certain $g$-factor distribution. Studies on Zeeman splitting in PL measurements have already revealed quite a variation in $g$-factors within TMDCs \cite{NatNano.10.503, NatPhys.11.148}, especially if localized states at defects are involved \cite{NatNano.10.503}. For a Lorentzian distribution of $g$-factors with a half-width-half-maximum value of $\Delta g$ around the mean value $g$, the inhomogeneous spin dephasing time is given by \cite{NanoLett.15.8250, PhysRevB.88.155316}:
\begin{equation}
\frac{1}{\tau_{\text{s}}^{\text{inh}}} = \frac{1}{\tau_{\text{s}}^{\text{0}}}+ \frac{\Delta g \mu_{\text{B}} B}{\hbar},
\label{eq:1overB}
\end{equation}
where $\tau_{\text{s}}^{\text{0}}$ is the spin lifetime at zero magnetic field. The fit of the inhomogeneous spin dephasing time $\tau_{\text{s}}^{\text{inh}}$ in Fig.~\ref{fig1}(c) with this formula yields $\Delta g = 0.33\pm 0.03$, which interestingly is equal to the mean value of the $g$-factor extracted by fitting the Kerr-data with Eq.~\ref{eq:threeExp} which gives $g \approx 0.33$. A previous study reported a sample-to-sample variation of the $\Delta g$ value in MoS$_2$ between $0.04<\Delta g<0.12$ \cite{NanoLett.15.8250} and further below we show that this value can change over time even within the same sample and, thus, is most likely influenced by extrinsic factors.

In the supplemental material we show that $\Delta g$ is a robust fitting parameter \cite{Supplement}, whereas there is quite an uncertainty regarding the value of $\tau_{\text{s}}^{\text{0}}$. In Fig.~\ref{fig1}(d) we plot Eq.~\ref{eq:1overB} with $\Delta g = 0.33$ for the assumption that the lifetime of the inhomogeneously dephasing spin component at zero magnetic field is equal to the magnetic-field independent, long-lived trion lifetime measured in TRR, i.e. $\tau_{\text{s}}^{\text{0}} = \tau_{\text{s}}^{\text{long}} = T_{\text{rec}}^{\text{long}} = \unit{150}{ns}$. This assumption yields an excellent agreement to the experimental data, which points to the fact that the magnetic field-independent long-lived trion states in TRR and the inhomogeneously dephasing spin component in TRKR have the same origin. Thus, equation~\ref{eq:threeExp} can be written as:
\begin{eqnarray}
\Theta_{\text{K}}&=& A_{\text{s}}^{\text{short}} \text{exp} \left( -\frac{t}{T_{\text{rec}}^{\text{short}}} \right)+ A_{\text{s}}^{\text{long}} \text{exp} \left( -\frac{t}{T_{\text{rec}}^{\text{long}}} \right) \nonumber \\
&+& \Bigg[ \sum\limits_n  A_{\text{s}}^{\text{inh}} \text{cos} \left( \frac{g \mu_{\text{B}} B}{\hbar} (t+n  T_{\text{rep}})  + \phi \right) \nonumber \\
&& \text{exp} \left( -(t+n  T_{\text{rep}}) \left[\frac{1}{T_{\text{rec}}^{\text{long}}}+ \frac{\Delta g \mu_{\text{B}} B}{\hbar} \right] \right) \Bigg],
\label{eq:inh}
\end{eqnarray}
which means that the whole spin dynamic observed in TRKR measurements can be described by excitonic recombination times, which are independently determined from TRR measurements, and the existence of a Lorentzian-shaped $g$-factor distribution. Here, the magnetic-field independent trion spin-component can be attributed to the hole spin of the trion, which cannot precess because of the large spin-orbit splitting in the valence bands \cite{PhysRevB.88.085433}, while we assign the electron spins of the trion states to the contribution which shows spin precession. The observation of $\tau_{\text{s}}^{\text{0}}=\tau_{\text{s}}^{\text{long}}=T_{\text{rec}}^{\text{long}}$ implies that the spin dynamics in TRKR is restricted to trion spin dynamics with the spin lifetimes limited by the trion recombination times and, hence, that there is no transfer of spin information to resident carriers in our sample, which is in contrast to the conclusions of several previous studies \cite{NatureComm.6.896,NanoLett.16.5010,NatPhys.11.830,NanoLett.15.8250,arXiv160203568B}. We note that lifetimes seen in our Kerr-measurements should not be mixed up with the one-particle spin lifetimes of free charge carriers. This is the reason why we denote the extracted lifetimes as excitonic or trion spin lifetimes (a more detailed discussion of our notation can be found in the supplemental material~\cite{Supplement}).

We note that Eq.~\ref{eq:threeExp} is not able to reliably extract $\tau_{\text{s}}^{\text{inh}}$ for magnetic fields smaller than $\unit{1}{mT}$ or larger than $\unit{100}{mT}$. This is due to the fact that for small magnetic fields both the lifetime of the precessing electron and non-precessing hole spin become comparably long which does not allow them to be distinguished by the fit routine. Whereas for magnetic fields above $\unit{100}{mT}$ the strongly dephased electron spins exhibit spin dephasing times which are comparable to the short-lived exciton lifetime. But as the decay time of the precessing electron spin is no longer an independent fit parameter in Eq.~\ref{eq:inh}, the Kerr rotation curves can now also be fitted for these magnetic fields. This is shown by the dashed red lines in Fig.~\ref{fig1}(c) for $B=\unit{0.68}{mT}$ and $B=\unit{100}{mT}$, which represents fits by Eq.~\ref{eq:inh}, whereas the red lines are fitted according to Eq.~\ref{eq:threeExp}. The corresponding values for $\tau_{\text{s}}^{\text{inh}}$ are marked as red triangles in Fig.~\ref{fig1}(d).

It is important to note that the measured lifetimes can vary between different cooling cycles and, hence, the data which we discuss next do not exceed the value of \unit{100}{ns} anymore. We attribute this change in lifetimes to extrinsic effects, as e.g. adsorbed molecules on top of 2D materials can have quite a significant impact on their properties \cite{ACSNano.10.3900}. In case of TMDCs it was shown that different gaseous environments change the PL-spectrum of monolayer flakes \cite{NanoLett.13.2831}. Thus, we assume that the change in lifetimes is linked to variations in the high-vacuum conditions of the sample chamber right before the next cool-down, which leads to different layers of condensed residual gases on top of the TMDC-flake at cryogenic temperatures. Another explanation could be laser-induced cleaning effects of the flake during the actual measurements as it was observed for graphene \cite{NanoMicroLetters.8.336}. However, despite the change in absolute values of the lifetimes, we find that the underlying spin dynamics do not change. In other words, Eq.~\ref{eq:inh} remains valid for each individual cooling cycle. This is demonstrated, e.g., in the next set of measurements where we recorded the TRR and TRKR curve at a fixed probe energy in the trion energy regime and vary the pump energy.

In Fig.~\ref{fig2}(a) we show the recombination lifetimes for $E_{\text{probe}}=\unit{1.715}{eV}$ at \unit{5}{K} and \unit{100}{mT} as a function of pump energy as green circles. As discussed above these values are equal to the respective trion spin lifetimes. In this figure also a close-up of the PL spectrum from Fig.~\ref{fig1}(b) is included demonstrating the position of exciton and trion states. As long as the pump energy is within the trion energy regime, the extracted lifetimes are more or less constant within the range of measurement uncertainty. We also measure a large Kerr rotation signal at trion energies when the pump energy is set to the neutral exciton energies (around $E_{\text{pump}}=\unit{1.746}{eV}$). This can be explained by an energetical decay of the neutral spin polarized exciton states into spin-polarized trion states by catching an additional electron. Apparently, the measured spin lifetime in the trion regime is reduced when pumping into neutral exciton states at higher energies. This may be explained by the fact that the excited bright neutral excitons not only decay into bright trion states (and these states to dark trion states) but they can also decay into dark neutral exciton states. The latter have an energy similar to the trions \cite{PhysRevLett.115.257403,PhysRevB.92.125431, PhysRevB.93.121107} and may lead to additional spin dephasing by many-body interaction between dark neutral and dark trion states.

The change in spin dynamics when pumping exciton states and probing at trion energies gets obvious in magnetic field dependent TRKR measurements which are shown for three different pump energies in Fig.~\ref{fig2}(b) and were fitted to Eq.~\ref{eq:inh} with $g\approx \Delta g=0.3-0.4$. The respective fit curves are included as red lines in Fig.~\ref{fig2}(b). The trion spin lifetimes of the long-lived Kerr-component are given by the triangular symbols in Fig.~\ref{fig2}(a) and again nicely match the recombination times. However, we note that the short-lived recombination time in the TRR measurements is not equal to the short spin lifetime in TRKR measurements when pumping higher energy neutral exciton states, which we discuss in the supplemental material \cite{Supplement}. In case of the two pump energies within the trion regime we remeasured the magnetic field TRKR data in a later cooling-cycle and observed an increase in the measured lifetimes (quadratic symbols, the blue one stems from the data set of Fig.~\ref{fig1}(c)).

To further explore the fine structure of the trion spin states we now focus on the energy dependence of the Kerr resonance. In Fig.~\ref{fig2}(c) we plot the Kerr rotation amplitude of the long-lived Kerr-signal as black squares. The resonance was recorded by sweeping the pump energy while the probe energy was fixed at $E_{\text{probe}}=\unit{1.715}{eV}$. The condition of equal pump and probe energy yields a clear extremum with negative sign in the amplitude of the Kerr rotation. The Kerr rotation changes sign and shows a positive maximum when pumping approximately \unit{7}{meV} below the probe energy. We next clarify the origin of the observed energy splitting of \unit{7}{meV} between the maximum and the minimum of the Kerr resonance curve and its dependence on both probe and pump energies. We interpret this splitting by the fine-structure of the trion state and note that n-type tungsten-based TMDCs exhibit two negatively charged bright trion states \cite{NatSciRev.2.57} (see Fig.~\ref{fig4}(b) and \ref{fig4}(c)). Both states differ in energy depending on whether the additional electron is in the same or in the opposite valley with respect to the valley in which the electron-hole pair has been initially excited. Because of the spin-split band structure of tungsten-based TMDCs, the two electrons of the negatively charged trion can thus have opposed spins (singlet trion in the intra-valley case) or equal spins (triplet trion in the inter-valley case). Due to exchange interaction, an energy splitting of \unit{6}{meV} is predicted between the singlet and triplet states \cite{NatureComm.5.3876}, which is in good agreement to the observed splitting of about \unit{7}{meV}. This value is also in perfect accordance to the measured trion state splitting in PL measurements on WSe$_2$ \cite{NatPhys.12.323}. A singlet-triplet splitting with a sign reversal of the respective Kerr rotation signals of the trion states has also been observed in WS$_2$ with a value of \unit{11}{meV} \cite{NatureComm.7.12715}.

As we use two independently tunable laser-systems, we repeated the pump scan of the Kerr rotation amplitude for different probe energies ($T=\unit{5}{K}$ and $B=\unit{0}{T}$) which is shown in Fig.~\ref{fig2}(d). Surprisingly, we observe a continuous shift of the Kerr resonance with probe energy. There is always a local maximum with negative sign for $E_{\text{pump}}=E_{\text{probe}}$ (this condition is indicated by black arrows, dashed lines represent $\Theta_K=0$) and another local maximum with inverted sign at a pump energy which is always \unit{7}{meV} below the probe energy (indicated by red arrows). This continuous shift will be explained further below by an adsorbate-induced inhomogeneous potential landscape of the flake, meaning that extrinsic effects have a significant influence to the overall measured spin-dynamics.

\begin{figure*}[tb]
	\includegraphics{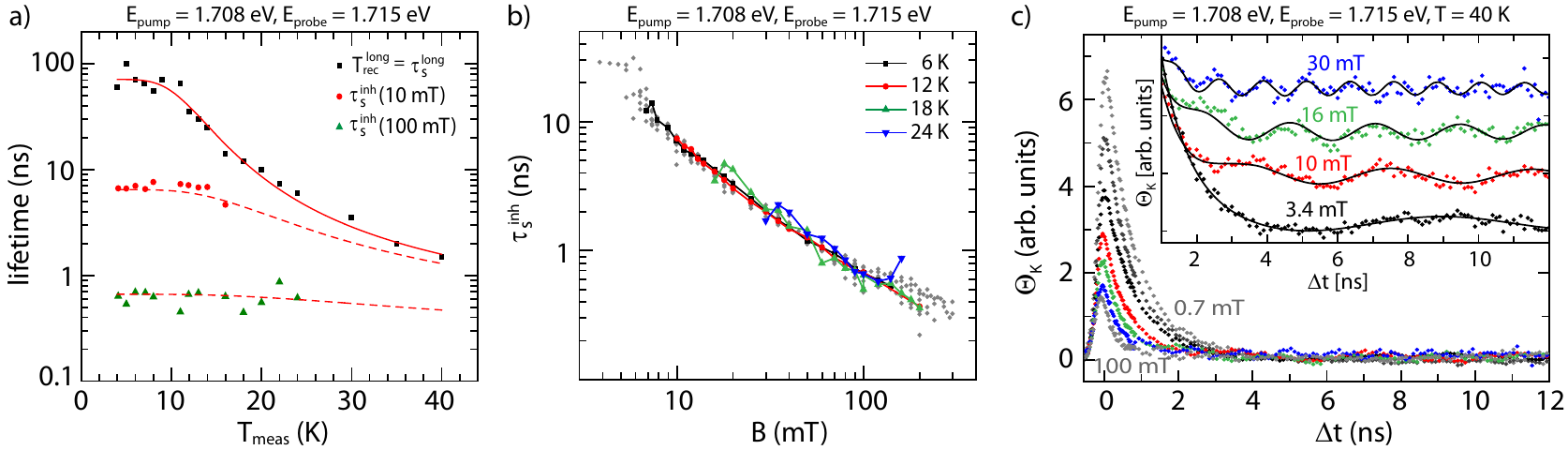}
	\caption{(a) Temperature dependent lifetimes extracted from Kerr and reflectivity measurements for the same probe/pump energies as in Fig.~\ref{fig1}. The long-lived recombination/excitonic spin-signal is fitted by assuming a thermally activated dephasing channel (equation \ref{eq:thermal} of the main text). The activation energy was set to the observed energy splitting of \unit{7}{meV} (see Figs.~\ref{fig2}(c) and \ref{fig2}(d)). (b) The inhomogeneous spin component shows a clear $1/B$-dependence up to temperatures of \unit{30}{K}. The gray points represent data from other temperatures between \unit{6}{K} and \unit{30}{K}. Interestingly, the inhomogeneous spin-component has no discernible temperature dependence as the measurements for different temperatures lie on top of each other. (c) At \unit{40}{K} the three different components, which can be observed at low temperatures (see e.g. Fig.~\ref{fig1}(c)), are no longer distinguishable from each other and the Kerr-signal can be described by a single exponential decay within the first \unit{2}{ns}. Interestingly, we observe a long-lived, oscillatory spin-signal ($g\approx 1.83-1.90$) with a small amplitude for longer timescales. The inset in (c) is a zoom-in of the tails with an vertical offset for clarity.}
	\label{fig3}
\end{figure*}

The important role of singlet and triplet states for the spin dynamics also becomes evident as their energy-splitting of \unit{7}{meV} is also observed in temperature dependent measurements. This is shown in Fig.~\ref{fig3}(a) where we plot the temperature dependent lifetimes extracted from TRR and from magnetic-field dependent TRKR measurements. The long-lived trion spin component, which once again was verified to be equal to the long-lived recombination signal, exhibits a small plateau below \unit{10}{K} and then undergoes a rapid temperature dependent dephasing. The long-lived trion spin lifetime can be fitted by assuming a thermally activated dephasing channel of the form (see e.g. \cite{NatPhys.11.477}):
\begin{equation}
	\frac{1}{\tau_{\text{s}}^{\text{long}}}=\frac{1}{\tau_0}+\gamma \; \text{exp}\left( \frac{-E_{\text{therm}}}{k_{\text{B}}(T_{\text{meas}}+T_{\text{off}})} \right),
\label{eq:thermal}
\end{equation}
with a temperature independent scattering rate $1/\tau_0$, which is responsible for the plateau at low temperatures, and a thermally activated scattering process with a proportionality factor $\gamma$ and an activation energy $E_{\text{therm}}$. Considering both the heating of the sample by the laser pulses and its finite thermal conductance to the cold-finger of the cryostat, there is most likely a slight difference between the actual sample temperature $T$ and the measured temperature $T_{\text{meas}}$ of the cold-finger, which we consider in the fitting by the additional parameter $T_{\text{off}}$. The red solid line in Fig.~\ref{fig3}(a) represents a fit, where the activation energy was set to the observed energy splitting of \unit{7}{meV} between singlet and triplet states. This fitting yields a thermally activated dephasing rate of $\gamma=\unit{4.3\cdot}{1/ns}$ and a sample temperature which is \unit{1.7}{K} above the measured one on the cold-finger. Although the fit catches the temperature dependence quite well, which is a strong indication that the singlet-triplet splitting is a key feature to understand the spin dynamics, we note that there is quite some leeway in fitting this data. As we show in Ref.~\cite{Supplement} the thermal activation energy can only be determined within a range of \unit{6-10}{meV}. Here, \unit{6}{meV} is the value when assuming that there is no temperature difference between the sample and the cold finger, which is nevertheless in good accordance to the energy splitting observed in the TRKR measurements and the prediction of Ref.~\cite{NatureComm.5.3876}.

In Fig.~\ref{fig3}(a) we plot the temperature dependent spin lifetime of the spin precessing component for two distinct magnetic fields. The dashed lines are calculated by putting the fitted temperature dependence of $\tau_{\text{s}}^{\text{long}}$ from Eq.~\ref{eq:thermal} into Eq.~\ref{eq:1overB}. The value for $\Delta g$ was determined with the same procedure as for the data in Fig.~\ref{fig1}, i.e. by fitting the whole set of magnetic field dependent TRKR measurements for each temperature with Eq.~\ref{eq:threeExp} (see Fig.~\ref{fig3}(b) for the resulting values of  $\tau_{\text{s}}^{\text{inh}}$) and fitting the magnetic field dependent spin lifetimes with the inhomogeneous dephasing model using Eq.~\ref{eq:1overB}. In Fig.~\ref{fig3}(b) the magnetic field dependent curves for each temperature lie on top of each other, which may suggest that there is no temperature dependence of the spin component at all. But we note that we limit the analysis to the case where the $B$-field dependent spin dephasing term in Eq.~\ref{eq:1overB} dominates the first, temperature dependent term ($\tau_{\text{s}}^{\text{0}} = \tau_{\text{s}}^{\text{long}}(T)$). Hence, the curves in Fig.~\ref{fig3}(b) start at larger $B$-fields when increasing the temperature (see also the suppressed temperature dependence at higher magnetic fields in case of the dashed lines in Fig.~\ref{fig3}(a)).

Although the trion spin lifetimes, the amplitudes, and the $g$-factor distributions changed again between cooling cycles, the underlying spin dynamic does not change and all equations discussed so far were applicable. The Lorentzian distribution of $g$-factors determined by Eq.~\ref{eq:threeExp} and Eq.~\ref{eq:1overB} now exhibits a mean value of $g \approx 0.20$ and a half-width-half-maximum value of $\Delta g \approx 0.17$, which is approximately a factor of two lower than in case of the very first measurements. This reduction in the distribution may be caused by desorption of adsorbates caused by the repeated pumping before each cool-down or to a more homogenous spatial distribution of the adsorbates due to laser-induced diffusion. Finally, we note that for temperatures above \unit{40}{K} the three distinctly different components which can be identified at low temperatures (i.e. $\tau_{\text{s}}^{\text{long}}, \tau_{\text{s}}^{\text{inh}}$, and $\tau_{\text{s}}^{\text{short}}$) are no longer distinguishable. Instead, the whole TRKR signal can be described by a single exponential decay within the first couple ns (Fig.~\ref{fig3}(c)). Interestingly, as soon as we enter into this regime, we observe a clear Larmor spin precession signal with $g\approx 1.83-1.90$ on longer timescales. This is shown in the inset of Fig.~\ref{fig3}(c), which depicts a close-up of the tails with an vertical offset for clarity. As the $g$-factor is significantly larger than the above g-factor of the trion states ($g \approx 0.20-0.33$), the oscillatory spin states are of different electronic origin which demonstrates the complexity of the overall spin dynamics.

We now concentrate on the explanation why adsorbate-assisted scattering of bright excitons into dark states can be responsible for the observed spin phenomena and the long spin lifetime which is orders of magnitude longer than the ones observed in other WSe$_2$-studies \cite{PhysRevB.90.161302,NatureComm.6.896,arXiv150704599Y}. Dark states seem to be a necessary requirement to reconcile the ps lifetimes reported in time-resolved PL measurements \cite{PhysRevB.90.075413, APL.105.101901,NatPhys.11.477, PhysRevB.93.205423, PhysRevLett.112.047401, PhysRevB.93.205423,Nanoscale.7.7402} with the ns lifetimes reported in this and in other studies \cite{NatPhys.11.830, NanoLett.15.8250, NanoLett.16.5010, arXiv160203568B}. This is because a PL measurement only detects the recombination of bright excitons, whereas TRR and TRKR measurements are sensitive to the imbalance in the population of charge and spin states. But such an imbalance can also be caused by dark states, i.e. many body-particles that cannot recombine by the emission of a photon and which are therefore not visible in PL measurements. Furthermore, we note that there is an interesting study which combines spatially-resolved Kerr measurements on a WS$_2$ monolayer with spatially-resolved PL measurements \cite{arXiv160203568B}. This study demonstrates that long lifetimes in TRKR measurements can be measured in regions with low PL intensity and vice versa. Furthermore, it is argued that the exchange coupling and the (pseudo-)spin relaxation is significantly suppressed in dark states in comparison to bright states \cite{2DMaterials.3.035009,PhysRevX.6.021024}. All this points to the fact that dark excitons might be of great importance to explain the spin-dynamics in TMDCs.

The notion that adsorbates play a significant role in the observed spin dynamics not only stems from the changes of the lifetimes over time but also from studies of a different 2D-material, namely graphene, where also varying spin lifetimes from ps to ns can be measured in electrical spin transport measurements \cite{Nature.448.571,NanoLett.16.3533}. In case of graphene extrinsic effects like substrate-, contact-, or impurity-induced spin dephasing mechanism are argued to be the reason for this large variation in the measured spin lifetimes \cite{PhysRevLett.112.116602,PhysRevB.90.165403,PhysRevB.80.041405}. In this context, we note that there is one TRKR study on WSe$_2$ which reports about spin lifetimes of several tens of ns \cite{NanoLett.16.5010}. As in our case, the PL spectrum in Ref.~\cite{NanoLett.16.5010} is dominated by localized states and exhibits an extremely small free exciton peak, which might be explained by a large, impurity- or defect-induced doping. But we also note an interesting difference between both studies, which sheds light on the underlying spin dynamic: Whereas Ref.~\cite{NanoLett.16.5010} reports on three, magnetic field independent components in their Kerr rotation signal, we only have two of such components and the third one exhibits the inhomogeneous dephasing mechanism. One possible explanation for this difference can be found in the fact that Ref.~\cite{NanoLett.16.5010} reports a p-doping of their sample, whereas we most likely have an n-doped monolayer which suggests that the spin component which exhibits inhomogeneous spin dephasing might be related to the second electron of the negatively charged trion.

Another reason for the long spin lifetimes might be that we probe spin states in the trion and not in the exciton regime as the inter-valley scattering rate, which was previously attributed to be the main channel for spin dephasing \cite{NatPhys.11.830, PhysRevB.92.235425,NatureComm.6.896}, is predicted to be much smaller for trions compared to excitons \cite{arXiv160800038S}. Interestingly, if inter-valley scattering dominates, a $1/B^2$-dependence of the spin lifetime is expected \cite{NatPhys.11.830} which contradicts the $1/B$ dependence in Fig.~\ref{fig1}(d) (a detailed discussion of this aspect can be found in the supplemental material \cite{Supplement}). Hence, in our sample inter-valley scattering does not seem to be of relevance for the observed trion spin lifetimes of up to \unit{150}{ns} and will be ignored in the following.

Furthermore, time-resolved transmission experiments in WSe$_2$ demonstrates a much longer polarization decay time for the singlet state compared to the triplet state \cite{arXiv160800038S}. Interestingly, the study in Ref.~\cite{arXiv160800038S} and also the study in Ref.~\cite{NanoLett.16.5010} which demonstrates spin lifetimes of tens of ns in WSe$_2$ uses laser pulses in the ps-range as we do. On the other hand, all studies which report ps spin lifetimes in WSe$_2$ used fs-laser pulses \cite{PhysRevB.90.161302,NatureComm.6.896,arXiv150704599Y}. Hence, a narrow spectral width of the laser pulses may be another necessary prerequisite for the observation of long-lived excitonic spin components. Pulses in the fs-range have a spectral width which is comparable to the energy splitting of singlet and triplet states of the trions, whereas our ps laser pulses have a spectral width of less than $\unit{1}{meV}$. The fact that the temperature dependence of the excitonic spin lifetime in Fig.~\ref{fig3}(a) can also be fitted by a thermal activation energy equal to the energy splitting between singlet and triplet states confirms the notion that the lifetime is strongly reduced if singlet and triplet states are simultaneously excited. This may either result from spectrally broad laser excitation or by thermal smearing. If this notion is correct, the long-lived excitonic spin component has to be linked to only one of the trion's states, which will be one important aspect of the following model.

\begin{figure*}[tb]
	\includegraphics{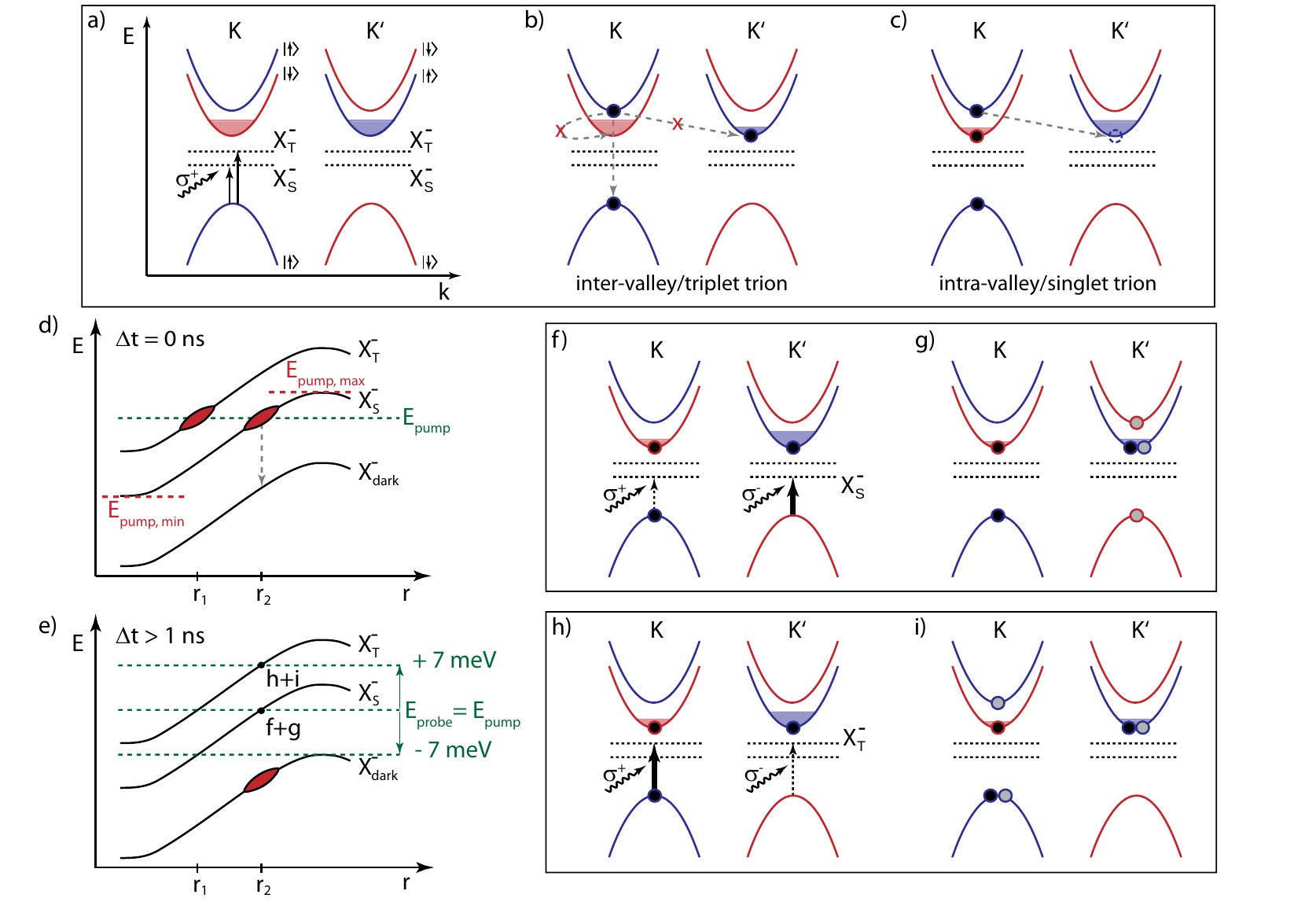}
	\caption{Illustration of the model that can explain both the observation of a Kerr-signal from the trion's higher-energy triplet state when the lower-energy singlet state is pumped (see Fig.~\ref{fig2}(c)) and the continuous shift in the Kerr-resonance with probe energy (see Fig.~\ref{fig2}(d)). See text for a detailed explanation. (a) Spin-resolved band structure of WSe$_2$ at the two K-points. The sample is assumed to be n-doped (shaded area in the lower conduction band). Depending on its energy, the right-circularly polarized pump beam can either excite the triplet $X_T^-$ (b) or singlet state $X_S^-$ (c). Only the singlet state is assumed to energetically relax into a long-lived, dark state through a scattering process of the upper conduction band electron into the K'-valley (c), whereas the triplet exciton (b) is assumed to recombine on a short time scale. (d) The sample exhibits a spatially varying potential landscape and, hence, a pump beam with a fixed energy will excite both singlet and triplet states, but at different locations (marked in red). (e) Only at the places where the singlet state was excited by the pump pulse a population of dark excitons remain. (f)-(i) These dark states block the formation of higher-energy, bright exciton states (both singlet and triplet states). As there is an asymmetrical population of the trion-bounded and free charge carriers within both K-valleys, this blocking mechanism is dependent both on energy and helicity of the probe pulse.}
	\label{fig4}
\end{figure*}

In our model we now incorporate all aspects discussed so far but also make some simplifications to prevent the model from becoming too complex. We will especially exclude complex many-body interactions which were observed in TMDCs before. These include, e.g., coherent exciton-trion coupling \cite{PhysRevLett.112.216804}, bandgap renormalization by optically-induced charge carriers \cite{NatPhoton.9.466}, strong inter-valley scattering by Coulomb interaction \cite{NanoLett.16.2945}, or the laser-induced Stark effect \cite{Science.346.1205}. We neglect these effects as they are only observed on rather short time scales in the fs to ps range, which is much shorter than our spin dynamics which extends over tens of ns. Furthermore, most of these effects require high fluences or, respectively, a large density of optically excited many-body particles. But we only use a fluence of about $\unit{1}{\mu Jcm^{-2}}$. With a typical optical absorption of the WSe$_2$ monolayer of $<10\%$ this fluence results in a trion density of $\unit{<4\cdot10^{11} }{cm^{-2}}$. At this density many of the complex many-body effects only play a minor role \cite{NatPhoton.9.466, NanoLett.16.2945, Science.346.1205}.

The more important assumption is that we neglect scattering processes which involve a spin flip. This includes the inter-valley scattering process between energy-degenerated K- and K'-valley bands with different spin orientations (see Fig.~\ref{fig4}(a)). Inter-valley scattering was previously claimed to be a main cause of spin dephasing \cite{NatPhys.11.830, PhysRevB.92.235425,NatureComm.6.896}, but it contradicts our magnetic-field dependent spin lifetimes (see \cite{Supplement}). Furthermore, the whole spin-dynamics in our sample can be described by the time constants of the recombination signal, which implies that the inter-valley spin scattering rate has to be much smaller than the recombination rate (i.e. that the inter-valley spin scattering lifetime is even much longer than \unit{150}{ns}). Finally, we note that the aim of our model is not to quantitatively explain the long trion spin lifetimes but rather to qualitatively explain the two maxima in the Kerr-rotation data, which are \unit{7}{meV} separated from each other and shift continuously with the probe energy.

Fig.~\ref{fig4}(a) shows the schematic spin-resolved band structure of WSe$_2$ at the two K-points \cite{PhysRevB.88.085433}. Within the same K-valley the lowest conduction band has a different spin-index than the highest valence band and the spin-indices reverse between K- and K'-valley. As we deduced an n-doping from the PL-spectrum, the lowest conduction band in each valley is partially filled with charge carriers, which is illustrated by the shaded red and blue areas. As the electron band structure describes single electron states and not many-body particles such as neutral excitons or trions, we adopt a representation commonly used for semiconductors: Whereas the energy of the many-body particle lies somewhere in the bandgap of the single particle picture (dashed lines for singlet and triplet states, respectively), we denote the origin of each single charge carrier of the trion by filled circles within the corresponding band structure. Depending on its energy, the right-circularly polarized pump beam ($\sigma^+$ polarization) can either excite a triplet $X_T^-$ (Fig.~\ref{fig4}(b)) or a singlet trion state $X_S^-$ (Fig.~\ref{fig4}(c)), which are expected to exhibit an energy splitting of \unit{7}{meV} \cite{NatureComm.5.3876, NatPhys.12.323}. Common to both trion states is the excited electron-hole pair in the K-valley, where optical selection rules demand that the electron is in the upper of the two depicted conduction bands \cite{PhysRevLett.108.196802, NatureComm.3.887}. The additional second electron of the trion stems either from the same valley (intra-valley or singlet trion in Fig.~\ref{fig4}(c)) or from the opposite valley (inter-valley or triplet trion in Fig.~\ref{fig4}(b)) which at the same time reduces the valley carrier densities in the respective conduction bands.

Now we incorporate different valley dynamics of the two trion states into our model which was also addressed in Refs. ~\cite{arXiv160800038S,NatPhys.12.323}. We assume that only the singlet trion state in Fig.~\ref{fig4}(c) can energetically relax into a long-lived, dark trion state through a scattering process from the upper conduction band in the K-valley into the K'-valley (see dashed line in Fig.~\ref{fig4}(c)) \cite{PhysRevB.94.075421}. The final dark trion state is also depicted in Fig.~\ref{fig4}(f). It is dark because an electron-hole recombination requires either a spin flip of the electron in the K-valley or a phonon-assisted scattering process of the electron in the K'-valley. We emphasize that the required change in momentum and energy for the transition from the bright singlet to the dark trion state can be mediated by scattering at defects or impurities. In this respect we recall the significant contribution of localized states to the overall PL-spectrum in Fig.~\ref{fig1}(b). This may hint to a significant concentration of such scattering centers. Therefore, the transition from the bright to the dark trion state can be quite efficient, as long as this scattering rate is much larger than the intra-valley singlet trion recombination rate. It is also important to note that the recombination rate of the bright singlet trion is predicted to decrease for higher doping densities \cite{PhysRevB.93.045407}. We thus conclude that a large part of the excited bright singlet trions scatters into long-lived spin polarized dark trion states. In contrast, this scattering process is highly diminished in the case of the triplet excitons, because the lower conduction band state in the K'-valley is already occupied by the trion's additional electron (Fig.~\ref{fig1}(b)). The only possibility for reaching the dark state in Fig.~\ref{fig1}(f) is by a spin flip scattering process from the higher into the lower conduction band of the K-valley. We assume that this scattering rate is much smaller than the triplet recombination rate and, hence, there is no significant occupation of dark trion states when pumping triplet trions.

These different scattering rates become important after we discuss the next aspect of our model which stems from the PL measurement. It is argued that local potentials arising from impurities during the fabrication process shift the local exciton binding energies and thus leads to an inhomogeneous energy broadening of PL peaks \cite{NatureComm.6.8315}. Consistent to this argumentation, it is already demonstrated that a thermal annealing step of a TMDC monolayer can lead both to a shift in energy and a smaller width of the corresponding PL peak \cite{NanoLett.13.2831}. Moreover, a clear separation between the singlet and triplet peak in a PL spectrum so far was only measured in case of a monolayer WSe$_2$ flake on top of hexagonal boron nitride \cite{NatPhys.12.323}. This is consistent to graphene, where substrate-induced fluctuations in the potential landscape are shown to be significantly reduced for graphene on boron nitride compared to graphene on SiO$_2$ \cite{NatMater.10.282}. Therefore, due to the wide PL peaks seen in Fig.~\ref{fig1}(b) we assume that our sample exhibits an inhomogeneous potential landscape which leads to spatially varying binding energies of the singlet $X_S^-$, triplet $X_T^-$, and dark $X_{\text{dark}}^-$ trion states (see Fig.~\ref{fig4}(d)). Hence, the pump pulse with a fixed energy will resonantly and simultaneously excite both triplet and singlet trion states, but at different locations on the flake (positions $r_1$ and $r_2$, respectively, the excited states are marked in red).

Now we consider the distinctly different scattering rates from both singlet and triplet trions into dark states. For time delays longer than $\Delta t = \unit{1}{ns}$ (i.e. longer than the fast decaying signal in the TRKR measurements) the excited triplet states are most likely optically recombined, whereas we assume that the singlet trion states scatter into the dark states as depicted in Fig.~\ref{fig4}(e) at position $r_2$. The only remaining charge and spin imbalance results from the dark trion state at this position. When probing these dark trion states by linearly polarized probe pulses which can be treated as a superposition of right ($\sigma^+$) and left ($\sigma^-$) circularly polarized light, there are two cases which have to be distinguished.

We first discuss the case where pump and probe energies are equal, which leads to the situation depicted in Fig.~\ref{fig4}(f). The singlet trions $X_S^-$, which were excited by the pump pulse in the K-valley, have bound a portion of the available electrons in the K-valley to form the trion states whereas the photo-excited electrons have scattered into the K'-valley. In the K'-valley the amount of charge carriers is already replenished, as all triplet trions at position $r_1$, which have bound electrons in this valley during excitation (see Fig.~\ref{fig4}(b)), are already recombined (see Fig.~\ref{fig4}(e)) for $\Delta t>\unit{1}{ns}$. But the number of available charge carriers is an important parameter in the formation of trions, as can be seen in gate-dependent PL measurements, where the trion states are significantly suppressed for decreasing doping levels \cite{NatNano.8.634, NatNano.9.268}. If we assume that the bottleneck for the excitation of trions is the amount of available charge carriers, then the $\sigma^-$ component of the linearly polarized light will be absorbed more strongly than the respective $\sigma^+$ component (depicted as differently thick arrows in Fig.~\ref{fig4}(f)), as the former can excite more singlet states in the K'-valley because of its higher concentration of available electrons. The trion, which gets primarily excited by the stronger absorption of the $\sigma^-$ component of the linearly polarized light, is depicted in Fig.~\ref{fig4}(g) as gray circles.

In the second case the probe pulse energy exceeds the pump pulse energy by \unit{7}{meV}. The initial situation is identical to the previous case, i.e. we are left with dark trion states at position $r_2$. But now the probe energy is in the range where its $\sigma^+$ and $\sigma^-$ components excite additional triplet trion states $X_T^-$ at position $r_2$. As a result the formation of trions depends on the number of electrons in the opposite valley with respect to the excited electron-hole pair. Therefore, in this case the $\sigma^+$ component of the linearly polarized light will now be absorbed more strongly than the respective $\sigma^-$ component (depicted as differently thick arrows in Fig.~\ref{fig4}(h)). Again, the trion which now gets primarily excited by the stronger absorption of the $\sigma^+$ component of the linearly polarized light is depicted in Fig.~\ref{fig4}(g) as gray circles.

Within this picture there is a change between the predominantly absorbed circularly polarized light components of the probe pulses between the cases of $E_{\text{probe}}=E_{\text{pump}}$ and $E_{\text{probe}}=E_{\text{pump}}+\unit{7}{meV}$. Now it is important to note that in Kerr geometry the rotation and ellipticity of the reflected probe pulse can exhibit complicated dependencies on differences in the index of refraction and the absorption for left and right circularly polarized light, which we discuss in more detail in Ref.~\cite{Supplement}. For the sake of simplicity, we assume that the complex refractive index makes the Kerr rotation sensitive to differences in absorption (a measurement of the respective Kerr ellipticity is included in Fig.~\ref{fig2}(c) which also exhibit a sign reversal and a peak-to-peak energy difference of \unit{7}{meV}). Under this assumption the model can explain the two peaks with different signs seen in the Kerr rotation amplitude in Fig.~\ref{fig2}(c). Moreover, the model can explain the continuous shift of this Kerr resonance with probe energy in Fig.~\ref{fig2}(d). This can easily be understood from Fig.~\ref{fig4}(d) where singlet trion states can be populated on different positions $r$ at any excitation energy between $E_{\text{pump, min}}$ and $E_{\text{pump, max}}$. The model will always predict two Kerr-signals with inverted signs if the probe energy is equal or \unit{7}{meV} above the pump energy.

Nevertheless, we note that our model cannot explain the whole Kerr-rotation dependence seen in Fig.~\ref{fig2}. E.g. for several probe energies there seems to be an additional feature in the Kerr signal, which shows up when the pump energy exceeds the probe energy by \unit{7}{meV} as shown in Fig.~\ref{fig2}(c) and indicated by green arrows in Fig.~\ref{fig2}(d). Within the model, this feature is most likely caused by the triplet states excited by the pump pulse at position $r_1$ of the flake. Because according to Fig.~\ref{fig4}(e) the probe pulses with an energy \unit{7}{meV} below the pump pulse energy will probe the singlet trion state at the same position $r_2$, where the pump pulses created triplet trions. Hence, the assumption that all triplet states optically recombine on a short timescale seems to be only a first order approximation.

Although a more sophisticated model might be necessary to explain the Kerr-signal in every detail, we nevertheless believe that the basic idea of our model might be quite fundamental to understand the spin-dynamics in TMDCs: Singlet and triplet states will show opposite dependencies on an asymmetric population of charge carriers or many-body exciton states between the K- and K'-valley, because their formation requires charge carriers from opposite valleys. Furthermore, the incorporation of dark trion states most likely is a necessity to reconcile the much longer trion and spin lifetimes observed in our study or the one in Ref.~\cite{NanoLett.16.5010} compared to the much shorter exciton lifetimes obtained from time-resolved PL measurements.

In conclusion, we presented a detailed study of time-resolved Kerr rotation and time-resolved reflectivity measurements on a monolayer WSe$_2$ flake, where the exact values of the trion spin lifetime and the measured distribution of $g$-factors undergo a change over laboratory time, which we attribute to adsorbate-induced effects. An adsorbate-induced, inhomogeneous potential landscape can also explain the continuous shift of the Kerr resonance with the probe energy. This demonstrates that disorder and extrinsic effects are important aspects in understanding the spin-dynamics in TMDCs. The long trion spin lifetimes of up to \unit{150}{ns} are explained by adsorbate-assisted scattering of bright trions into spin polarized dark trion states. The absence of an impurity-induced relaxation channel may explain the significantly lower spin lifetimes typically observed in other studies \cite{Supplement}. Furthermore, we expect that this finding may open the way for a systematic tailoring of the spin properties of transition metal dichalcogenide monolayers. We also explored the singlet-triplet fine-structure of the spin-polarized trion states and observed a spin signal from the trion's higher-energy triplet state when optically pumping into lower-energy singlet states which we explain by the blocking of the latter states by lower-energetic dark trion states.

The research leading to these results has received funding from the European Union Seventh Framework Programme under grant agreement n${^\circ}$604391 Graphene Flagship.

\clearpage
\newpage

\renewcommand{\topfraction}{1.0} 
\renewcommand{\bottomfraction}{1.0} 
\renewcommand{\floatpagefraction}{1.0} 

\textbf{Supplement: Inter-valley dark trion states with spin lifetimes of 150 ns in WSe$_2$}

\subsection{Trion/excitonic spin lifetime}

In this Supplemental Material we first want to note that the decay times observed in time-resolved Kerr rotation or time-resolved Faraday rotation measurements can be caused and limited by quite different physical mechanisms. This is most likely the reason for the large variations in reported spin lifetimes of TMDCs so far (see table at the end of this Supplemental Material). E.g., in our sample the recombination time of the excitonic states is the limiting factor for the decay time of the TRKR signal. On the other hand, there are also studies showing decay times of the TRKR signal much longer than the measured recombination time and hence the recorded spin signal is dominated by different recombination/scattering mechanisms of particles other than excitons (e.g. by inter-valley scattering of spin-polarized resident charge carriers).

In this respect, we note that the term ``spin lifetime'' is quite a general term which can be applied to every observed decay constant in a time-resolved Kerr or Faraday measurement, regardless if the spin signal is caused e.g. by excitonic states or resident charge carriers and limited e.g. by recombination or inter-valley scattering. Unfortunately, this may lead to misunderstandings when comparing different studies and therefore we want to clarify our name conventions at this point.

\begin{figure}[b]
	\includegraphics{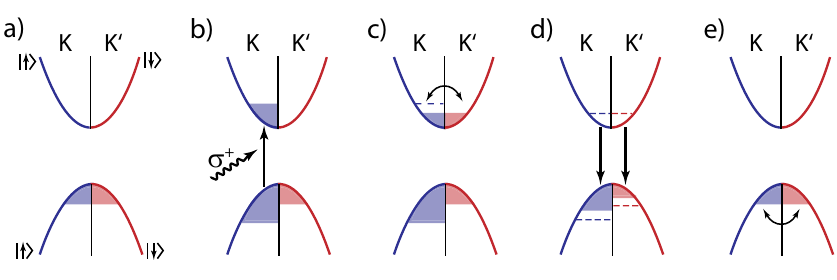}
	\caption{Illustrations for the name convention between spin lifetime vs. trion/excitonic spin lifetime. See text for a detailed explanation.}
\label{spin-polarization}
\end{figure}

To discuss this matter, we adopt a more simple picture compared to the one in the main manuscript and ignore the fact that many-body particles occupy other states than the ones in the single particle band structure. Furthermore, we limit ourselves to the excitation of neutral excitons and to a p-doped sample (see e.g. Refs.~\cite{NatureComm.6.896, NanoLett.16.5010}). The resulting schematic pictures are shown in Fig.~\ref{spin-polarization} of this Supplemental Material.

In Fig.~\ref{spin-polarization}(a) the steady-case situation is depicted, where the free holes in the valence band are represented by shaded areas. In Fig.~\ref{spin-polarization}(b) a right-circularly polarized pump beam excites excitons in the K-valley which, for the sake of simplicity, are depicted as a larger hole concentration in the valence band and an electron concentration in the conduction band of the K-valley. Such a situation will already lead to a Kerr or Faraday rotation signal as a linearly polarized probe pulse will experience different absorptions for its left- and right-polarized components, as both interact with different valleys. Therefore, in literature there is the argumentation that the pump pulse already generates a ``valley polarization''. In our opinion this can be misleading as long as we only consider the number of charge carriers in the respective valleys. E.g., in early publications a ``valley polarization'' is defined as different ``number of electrons in these valleys'' (see e.g. \cite{NatNano.7.490}). But the pump pulse ``only'' excites electrons from lower to higher energy states within the same valley and therefore the total number of electrons within each valley is still the same and especially identical to the other valley.

Of course, if we only consider the conduction band (or, respectively, only the valence band) there is an imbalance in charge carriers between both valleys. Hence, there is a ``conduction/valence band valley polarization''. But if both conduction and valence band are considered simultaneously, there is - strictly speaking - no (charge carrier) valley polarization. We note that there is also another way to define a (charge carrier) valley polarization, namely that a valley polarization is connected to different chemical potentials between both valleys \cite{PhysRevLett.99.236809}. Here, the problem lies within the definition of the chemical potential or Fermi energy in optical experiments, because the generation of excitons leads to a non-equilibrium condition.

Nevertheless, even in the case of Fig.~\ref{spin-polarization}(b) there exists a valley polarization if other quantities next to charge carriers are considered. E.g., the photons transfer their angular momentum into the K-valley by exciting electrons from the valence band to states with higher orbital angular momentum in the conduction band. Therefore, there is a polarization of the magnetic moment between both valleys. Another type of ``valley polarization'' in Fig.~\ref{spin-polarization}(b) is the imbalance in the number of excitons between both valleys, meaning that we have an ``excitonic valley polarization'' in our sample.

The situation shown in Fig.~\ref{spin-polarization}(b) can now develop in different ways. One possibility is that the excitons recombine, which directly leads back to the situation of Fig.~\ref{spin-polarization}(a). If this is the only process involved, then the lifetime in the TRKE signal should be identical to the recombination time of the TRR signal. This is the case in our measurements, with the only exception that we argue that there is an additional scattering into long-lived dark trion states before recombination occurs. Therefore, the observed spin lifetime in our TRKR signal may be described as an ``excitonic valley population spin lifetime'', or more simplified ``excitonic spin lifetime''.

We note that the term ``excitonic spin lifetime'' also may be applied to a situation where the exciton valley polarization gets diminished by inter-valley scattering, i.e. that after a certain time the amount of excitons in both K and K' valley is the same, although exciton recombination may not occur at this point. Such a situation may be described by some kind of ``excitonic inter-valley scattering spin lifetime''.

The situation of inter-valley scattering of only one of the two exciton charges is depicted in Fig.~\ref{spin-polarization}(c). If this process takes place before the actual exciton recombination (Fig.~\ref{spin-polarization}(d)), both a charge carrier valley and a spin polarization are present, as there are different amounts of spin-polarized holes in each valley and there exists a net spin polarization if all electrons in the system are taken into account. A basic requirement for the existence of this situation is a shorter recombination time compared to the lifetime extracted from TRKR measurements. In this case we may probe the ``(free) charge carrier spin lifetime'' of the TMDC, over which a spin polarization of free charge carriers relaxes into equilibrium (Fig.~\ref{spin-polarization}(e)).

Therefore, we once again emphasize that our measurements give no clues about the one-particle spin lifetimes of free charge carriers in WSe$_2$. As the lifetimes extracted from TRKR measurements are identical to the recombination times from the TRR measurements, there are two possible conclusions: Either there is no inter-valley scattering in our sample, which is a necessary requirement for the creation of an one-particle spin polarization of resident charge carriers (Figs.~\ref{spin-polarization}(c) and (d)). Or the one-particle spin lifetime is much smaller than the excitonic recombination time and therefore is not discernible in our experiments.

\subsection{Kerr-rotation and ellipticity}
The spin-related signal in magneto-optical measurements is caused by different complex refractive indices $\mathfrak{n}$ for right- (+) and left-circularly (-) polarized light, i.e. $\mathfrak{n_{\pm}}=n_{\pm}+i \kappa_{\pm}$ with the refractive index $n_{\pm}$ and the extinction coefficient $\kappa_{\pm}$ for each helicity. We define $\Delta n = n_{+}- n_{-}$ and $\Delta \kappa = \kappa_{+}- \kappa_{-}$ as the splittings between the right and the left-circularly polarized components and $n = (n_{+}+ n_{-})/2$ and $\kappa = (\kappa_{+}+\kappa_{-})/2$ as the mean real and imaginary parts.

It can be shown that in case of a Faraday geometry, the Faraday rotation is given by:
\begin{equation*}
 \Theta_F = \Re \left\{ \frac{\pi d}{\lambda} (\mathfrak{n}_{+}-\mathfrak{n}_{-}) \right\}= \frac{\pi d}{\lambda} \Delta n
\end{equation*}
and the ellipticity by
\begin{equation*}
 \eta_F = -\text{tanh} \left[ \Im \left\{ \frac{\pi d}{\lambda} (\mathfrak{n}_{+}-\mathfrak{n}_{-}) \right\} \right]= -\text{tanh} \left[ \frac{\pi d}{\lambda} \Delta \kappa   \right]
\end{equation*}
with the thickness $d$ of the medium and the wavelength $\lambda$ of the light \cite{PhysRev.97.334, IEEETransMag.4.152}. Therefore, the Faraday rotation only probes the difference in the real part of the refractive index $\Delta n$ and the Faraday ellipticity only probes the difference $\Delta \kappa$ of the extinction coefficient for right- (+) and left-circularly (-) polarized light.

In case of Kerr measurements, the dependency of rotation and ellipticity on the complex refractive indexes gets more complicated. First of all we note that the probe beam in our setup has an incident angle almost perpendicular to the sample surface and hence is almost parallel to the pump-induced spin polarization (the direction of the pump beam is perpendicular to the sample surface as shown in Fig.~\ref{setup}). There is only a small derivation of a few degrees of the perpendicular incidence in case of the probe beam, so that the reflected pulse can be separated from the other beams. Hence, we consider only the polar Kerr effect, where the Kerr rotation in first order is given by \cite{PhysRev.97.334, IEEETransMag.4.152}:
\begin{equation*}
\Theta_K \approx \Im \left\{ \frac{\mathfrak{n}_{+}-\mathfrak{n}_{-}}{\mathfrak{n}_{+} \mathfrak{n}_{-}-1} \right\}
\end{equation*}
and the Kerr ellipticity by
\begin{equation*}
\eta_K \approx \Re \left\{ \frac{\mathfrak{n_{+}}-\mathfrak{n_{-}}}{\mathfrak{n_{+}} \mathfrak{n_{-}}-1} \right\}.
\end{equation*}
In contrast to the Faraday rotation, the Kerr rotation not only probes $\Delta n$ but has a complicated dependence on both $\Delta n$ and $\Delta \kappa$.

We can replace $n_{\pm}$ and $\kappa_{\pm}$ in the above equations of $\Theta_K$ and $\eta_K$ by $n_{\pm}=(2 n \pm \Delta n)/2$ and $\kappa_{\pm}=(2 \kappa \pm \Delta \kappa)/2$ and solve the equations by ignoring higher-order terms of $\Delta n$ and $\Delta \kappa$, which then yields functions of the form:
\begin{eqnarray*}
\Theta_K&=&\alpha (n,\kappa) \Delta n + \beta (n,\kappa) \Delta \kappa \\
\eta_K&=&\gamma (n,\kappa) \Delta n + \delta (n,\kappa) \Delta \kappa,
\end{eqnarray*}
where the dependence of rotation and ellipticity is given as a function of the helicity dependent differences $\Delta n$ and $\Delta \kappa$ with proportionality factors $\alpha$ to $\delta$ depending on the median refractive index $n$ and extinction coefficient $\kappa$. Values of $n$ and $\kappa$ can e.g. be found in Ref.~\cite{APL.105.201905} for different monolayers of TMDCs. Putting these values in the above approximation in fact results in a Kerr rotation that highly depends on $\Delta \kappa$, which is a fundamental assumption of our model in the main manuscript.

However, we refrain to give exact numbers on the proportionality factors $\alpha$ to $\delta$, because the treatment is simplified. As we investigate a very thin film on a reflective substrate, we have to incorporated substrate-induced effects on the Kerr rotation signal, which makes the dependence of Kerr rotation and Kerr ellipticity on the complex refractive indices of all involved materials even more complicated \cite{JApplPhy.38.1652,CzechJPhysB.36.834,JMagnMat.89.107,JApplPhys.84.541}. A detailed analysis of the Kerr rotation and Kerr ellipticity as a function of $\Delta n$ and $\Delta \kappa$ by far exceeds the scoop of this paper. The only point we want to make is that the Kerr rotation is more complicated to analyze than the Faraday rotation and that the Kerr rotation very likely depend on $\Delta \kappa$ in accordance to our model.

\subsection{$1/B$-dependence and inter-valley scattering}
Here we want to discuss further conclusions which can be drawn from the observed $1/B$-dependence of the magnetic-field dependent spin component seen in the main manuscript. We note that the $1/B$-dependence of this component contradicts to a formula which was derived for the case that inter-valley scattering dominates the spin lifetime \cite{NatPhys.11.830}. This may be a very important observation as previous studies concluded that inter-valley scattering is the main cause of spin dephasing \cite{NatPhys.11.830, PhysRevB.92.235425,NatureComm.6.896}. Therefore, the fact that we measure spin lifetimes exceeding \unit{100}{ns} may be interpreted in such a way, that inter-valley scattering is somehow suppressed for the long-lived spin components observed in our sample.

In the ''Supplementary Information'' part of Ref.~\cite{NatPhys.11.830} the calculated formula for the spin-dynamic in case of small magnetic fields is:
\begin{equation*}
\Theta_{K} \propto \text{exp}\Big(-(\gamma_s+\Gamma_v)t \Big)   \left( \text{cosh}(\Omega t)  + \frac{\Gamma_v \text{sinh}(\Omega t)}{\Omega} \right)
\end{equation*}
with the intra-valley spin scattering rate $\gamma_s$, the spin conserving inter-valley scattering rate $\gamma_v$, $\Gamma_v = \Omega_{SO}^2/(4 \gamma_v)$ with the spin-orbit coupling $\Omega_{SO}$, and $\Omega = \sqrt{\Gamma_v^2-\Omega_L^2}$ with the Larmor frequency $\Omega_L = g \mu_B B / \hbar$.

\begin{figure*}[tb]
	\includegraphics{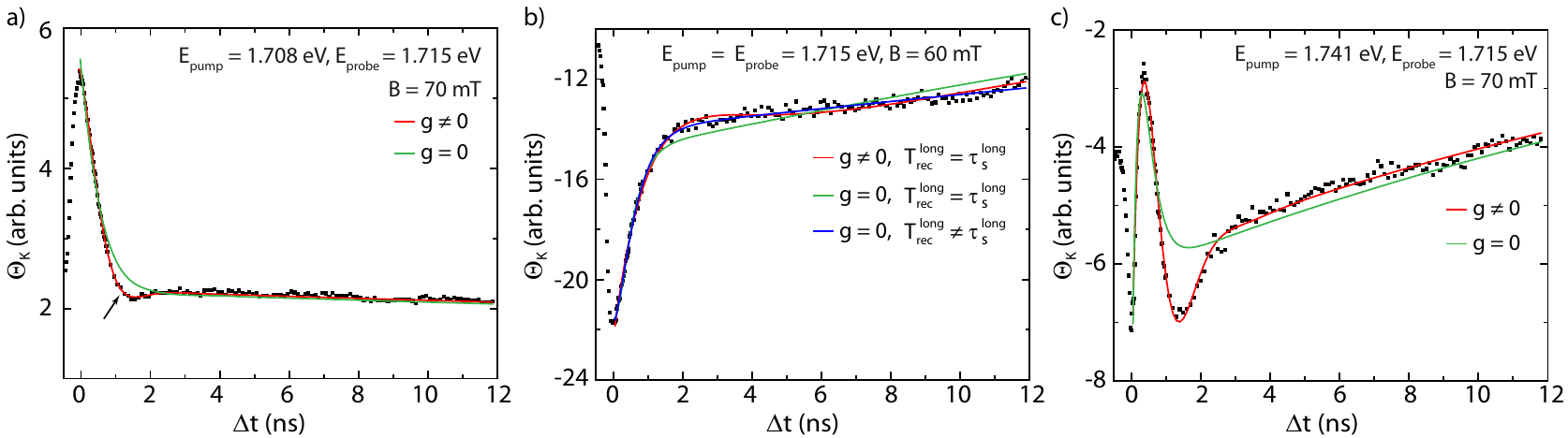}
	\caption{Fitting results to TRKR curves using equation 2 of the main manuscript ($g$-factor is a free fitting parameter, red curves) and fits with the same fitting function where the $g$-factor is set to zero (green curves). (a) For the majority of magnetic field dependent TRKR measurements a non-zero $g$-factor is needed to describe the small overshoot in the curves around a time delay of \unit{1-2}{ns} (indicated by an arrow). (b) In principal, a few curves can be fitted without a $g$-factor, but in these cases a good agreement between data and fit can only be achieved by dropping the assumption that the long-lived component of the recombination is identical with the long-lived component of the Kerr-signal (difference between green and blue curve). On the other hand, the assumption of a $g$-factor and identical recombination and excitonic spin lifetimes (red curve) can fit every single curve, which gives a far more consistent picture. (c) Especially if the amplitudes of the short-lived component and the one exhibiting  inhomogeneous spin dephasing have different signs, the assumption of a $g$-factor is crucial to fit the data.}
\label{g-factor}
\end{figure*}

Now we make the approximation $\Gamma_v/\Omega = \Gamma_v/ \sqrt{\Gamma_v^2-\Omega_L^2}\approx 1$ for small magnetic fields and use the Taylor-expansion $\sqrt{\Gamma_v^2-\Omega_L^2}\approx \Gamma_v - \Omega_L^2/(2\Gamma_v)$:

\begin{eqnarray*}
\Theta_{K} &\propto&\text{exp}\Big(-(\gamma_s+\Gamma_v)\;t \Big)   \left[ \text{cosh}(\Omega \;t)  + \frac{\Gamma_v \text{sinh}(\Omega \;t)}{\Omega} \right] \\
&\approx &\text{exp}\Big(-(\gamma_s+\Gamma_v)\;t \Big)   \Big[ \text{cosh}(\Omega t)  + \text{sinh}(\Omega t) \Big] \\
&=& \text{exp}\Big(-(\gamma_s+\Gamma_v)\;t \Big) \;  \text{exp}\left( \Omega t \right) \\
&=& \text{exp}\left(-\gamma_s \;t-\left[\Gamma_v- \sqrt{\Gamma_v^2-\Omega_L^2}\right]\;t \right) \\
&\approx& \text{exp}\left(-\gamma_s \;t-\left[\Gamma_v- \Gamma_v+\frac{\Omega_L^2}{2 \Gamma_v}\right]\;t \right) \\
&=& \text{exp}\left(-(\gamma_s+\frac{\Omega_L^2}{2 \Gamma_v})\;t \right).
\end{eqnarray*}
Hence, the magnetic field dependent spin component is (in first order) a mono-exponential decay function with an effective decay time of $1/(\gamma_s+\Omega_L^2/(2 \Gamma_v))$. The fact that it can be described by a single exponential decay function is in reasonable agreement to our measurements, as the additional spin precession term in our case (the cosine-term in equation 3 of the main manuscript) is in most cases only necessary to describe smaller features in the Kerr data, which is shown in Fig.~\ref{g-factor} of this Supplemental Material. But as $\Omega_L\propto B$, the magnetic-field dependent lifetime in case of inter-valley scattering should show a $1/B^2$-dependence, which is at odds with the $1/B$-dependence in our sample.

\subsection{Assumption in fitting the Kerr and reflectivity data}
As we described in the main manuscript, the overall reflectivity signal can be fitted by a bi-exponential decay function:
\begin{equation*}
	\frac{\Delta R}{R} = A_{\text{rec}}^{\text{short}} \text{exp} \left( -\frac{t}{T_{\text{rec}}^{\text{short}}} \right) + A_{\text{rec}}^{\text{long}} \text{exp} \left( -\frac{t}{T_{\text{rec}}^{\text{long}}} \right)
	\label{eq:rec}
\end{equation*}
with each a long and short recombination time $T_{\text{rec}}^{\text{i}}$ and amplitude $A_{\text{rec}}^{\text{i}}$. Here, we note one important assumption we made for this equation. Considering the fitting of the reflectivity data we have to take into account that the long-lived recombination time is much longer than the laser repetition time $T_{\text{rep}}$. Therefore, the influence of a new pump pulse on already existing population, which stem from previous pulses, has to be considered. This may be relevant as effects like optically-induced bandgap renormalization are already shown to be very strong in TMDCs compared to conventional semiconductors \cite{NatPhoton.9.466} and such effects may change the excitonic decay dynamics (and therefore the reflectivity signal) during the duration of each new incoming pump pulse. Thus, the overall reflectivity signal has to be fitted over the sum of several pulses:

\begin{eqnarray*}
	\frac{\Delta R}{R}&= &A_{\text{rec}}^{\text{short}} \text{exp} \left( -\frac{t}{T_{\text{rec}}^{\text{short}}} \right) \nonumber \\
	&+& \sum\limits_n \alpha (n, T_{\text{rec}}, \Phi, t) \tilde{A}_{\text{rec}}^{\text{long}} \text{exp} \left( -\frac{t+n  T_{\text{rep}}}{T_{\text{rec}}^{\text{long}}} \right),
	\label{eq:rec2}
\end{eqnarray*}
where the initially created amplitude $\tilde{A}$ of a pump pulse not only undergoes an exponential decay with a time constant $T_{\text{rec}}^{\text{long}}$ but also a modification by each new pump pulse, which is considered by an additional factor $\alpha$. Such a correction factor and its impact on TRKR measurements was already discussed e.g. in \cite{PSSB:PSSB201350201}. As effects like bandgap renormalization \cite{NatPhoton.9.466} or the laser-induced Stark effect \cite{Science.346.1205} significantly depend on the fluence $\Phi$, also the correction factor will depend on the fluence of the laser pulses. But as the fluence in our measurements is quite low (around $\unit{1}{\mu Jcm^{-2}}$), we neglect such effects and assume $\alpha (n, T_{\text{rec}}, \Phi)=1$, which leads to the first equation with $A_{\text{rec}}^{\text{long}} = \sum\limits_n  \tilde{A}_{\text{rec}}^{\text{long}} \text{exp} \left( -\frac{n  T_{\text{rep}}}{T_{\text{rec}}^{\text{long}}} \right)$.

\subsection{Robustness of the singlet/triplet splitting}
Here, we once again note that extrinsic effects, which were discussed in the main manuscript, can lead to changes in both lifetime and amplitude between nominally identical measurements in different cooling-cycles. E.g. in the main manuscript this can be seen in the difference between the curve in Fig.~2(c) and the one for the same probe energy in Fig.~2(d). On the other hand, both measurements were also performed at different magnetic fields. In case of Fig.~2(c) the applied \unit{100}{mT} is large enough to completely suppress the spin precession component of the electron spin, whereas the data set in Fig.~2(d) was measured at \unit{0}{mT}, where the spin precession component contributes maximal to the overall Kerr amplitude. But both data-sets show the same energy splitting of \unit{7}{meV}, meaning that the separation of singlet and triplet states is robust against both magnetic fields and temporal changes of the lifetime and amplitude of the Kerr signal.

\begin{figure*}[tb]
	\includegraphics{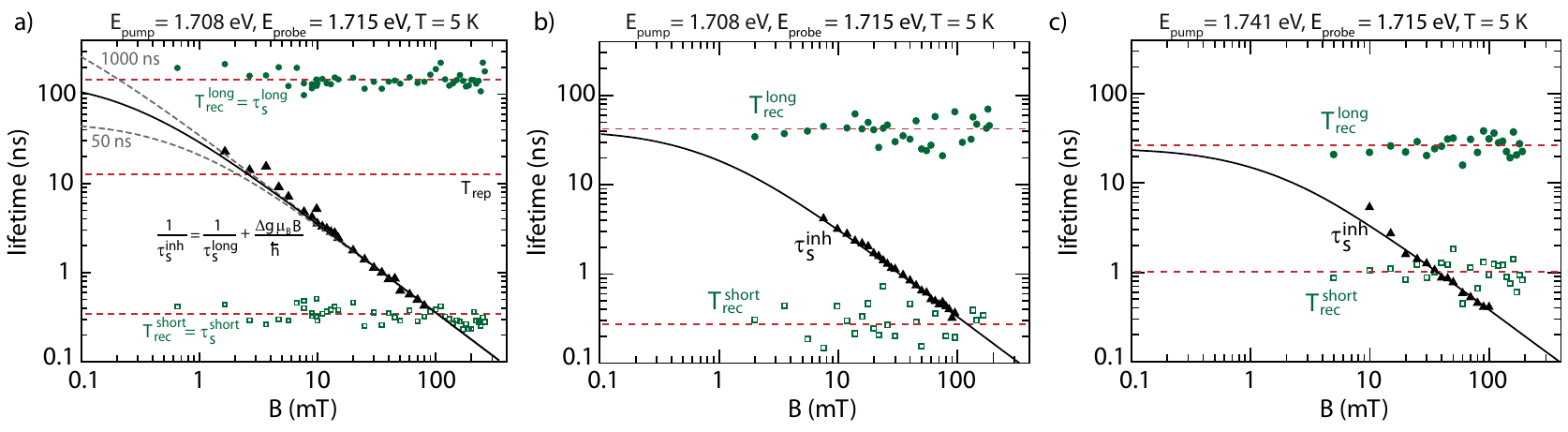}
	\caption{(a) and (b): Analysis of recombination times and inhomogeneous spin dephasing times for the same pump/probe energies at two different cooling circles. Figure (a) shows the same data as figure 1(d) of the main manuscript, while (b) corresponds to the set of curves in the upper panel of figure 2(b) of the main manuscript and was recorded one week before the data in (a). (c) The assumption that the short recombination time is identical to the short Kerr-signal is not valid in case of a pump energy in the exciton regime. Here the short recombination component is around \unit{1}{ns}, which is a factor of three larger than for pump energies in the trion regime.}
	\label{Kerr-rec}
\end{figure*}

\subsection{Differences between recombination and spin lifetime}
In the main manuscript we have shown that for pump energies in the trion regime both the long- and short-lived spin lifetimes are identical to the long- and short-lived recombination time, respectively. However, this is not true for a pump energy within the exciton range ($E_{\text{pump}}=\unit{1.741}{eV}$, green curves in Fig.~2(b) of the main manuscript and green triangle in Fig.~2(a) of the main manuscript). In this case only $\tau_{\text{s}}^{\text{long}}=T_{\text{rec}}^{\text{long}}$ is valid, whereas the short-lived recombination time is no longer identical to the short-lived Kerr-signal.

The short-lived component in the TRR signal in case of $E_{\text{pump}}=\unit{1.741}{eV}$ is a factor of three longer compared to other pump energies ($\unit{\approx1}{ns}$ instead of $\unit{\approx300}{ps}$, see Fig.~\ref{Kerr-rec}(c) of this Supplemental Material). However, the whole data set for this pump energy can be successfully fitted (Fig.~2(b) of the main manuscript) by assuming values of the short spin lifetime within the range that was observed for other pump energies, i.e. $\tau_{\text{s}}^{\text{short}}=\unit{\approx300}{ps}$. Accordingly, the inhomogeneous spin dephasing time can also be fitted if its value is below the one of the short recombination time.

The difference between short-lived recombination and excitonic spin lifetime may be explained by the fact that the pump energy is in the exciton regime, because next to trions the created exciton can also decay into dark excitons (also called indirect excitons). It may be important that the latter have an energy close to the one of trions \cite{PhysRevLett.115.257403,PhysRevB.92.125431}. Hence, the Kerr rotation signal may only probe the trion states, while the reflectivity, which we use as an indicator for the recombination, may also be influenced by dark excitons or higher-energy bright excitons.

\subsection{Uncertainty in $\tau_{\text{s}}^{\text{0}}$ and RSA measurements}
As mentioned in the main manuscript, there is quite an uncertainty regarding the value of $\tau_{\text{s}}^{\text{0}}$. In Fig.~\ref{Kerr-rec}(a) of this Supplemental Material we show the same data-set as in Fig.~1(d) of the main manuscript, but we included the dashed gray lines, which are the calculated curves for the same $\Delta g$ value but shorter ($\tau_{\text{s}}^{\text{0}}=\unit{50}{ns}$) or much longer ($\tau_{\text{s}}^{\text{0}}=\unit{1000}{ns}$) lifetimes, respectively. We note that the impact of different $\tau_{\text{s}}^{\text{0}}$-values only gets significant for magnetic fields smaller than $\unit{3}{mT}$. But in this range both the lifetimes of the inhomogeneously dephasing spin component and the long-lived magnetic field independent spin component are longer than the measured time delay. Therefore, they cannot be reliably resolved. In such a case, where the spin lifetime is much longer than the measurable time delay, normally the magnetic field dependent effect of resonant spin amplification (RSA) is used in order to reliably determine the spin lifetime \cite{PhysRevLett.80.4313}. In this kind of measurement a homogeneously precessing spin ensemble constructively interferes with the subsequent spin ensemble of the next laser pulse once the Larmor frequency matches a multiple of the laser repetition rate. But like we show in Fig.~\ref{helicity-RSA} of this Supplemental Material, the RSA-measurements on our sample only exhibit a single peak around $B=\unit{0}{mT}$. This means that the magnetic field dependent spin signal already dephased at magnetic field strengths smaller than the one required for the first resonant condition.

\begin{figure*}[b]
	\includegraphics{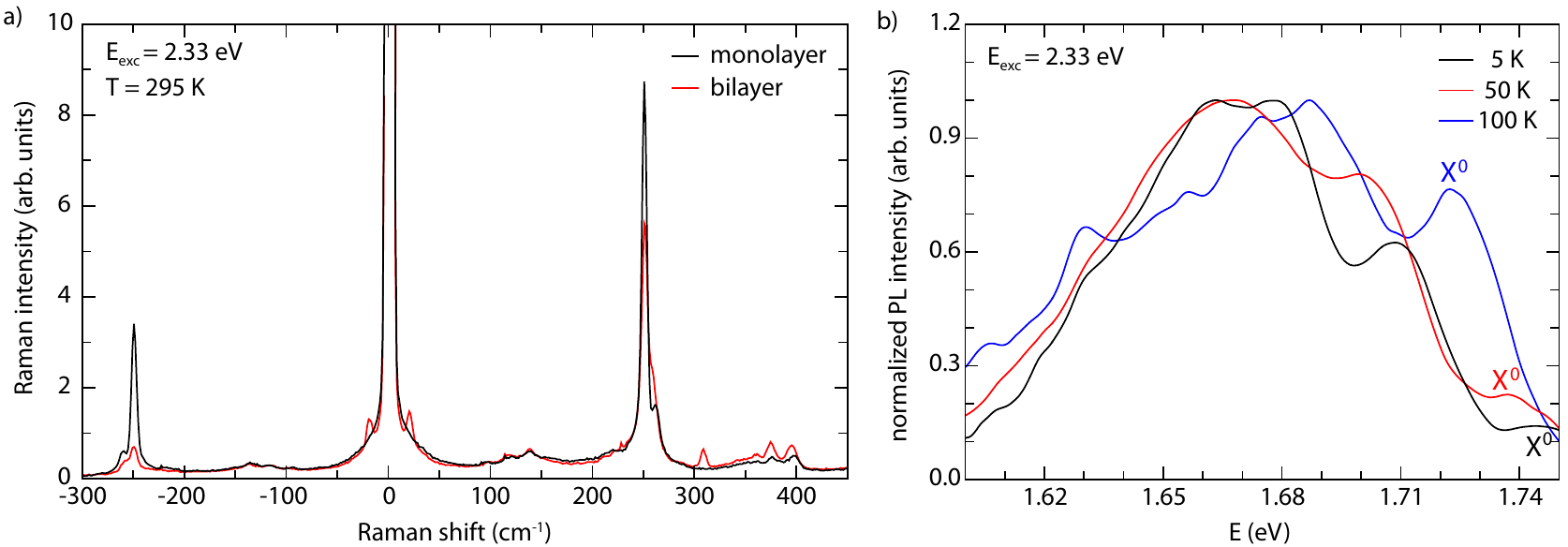}
	\caption{(a) Raman spectra of the monolayer (black curve) and bilayer (red curve) part of the WSe$_2$ flake at room temperature with a laser excitation energy of $E=2.33$~eV. The monolayer part was identified by the absence of the interlayer breathing and shear modes around $\pm$\unit{20}{cm^{-1}} \cite{NanoLett.13.1007,Chem.Soc.Rev.44.2757}. The dominant feature is the double peak between \unit{250}{cm^{-1}} and \unit{261}{cm^{-1}} which belongs to the $E'$ and $A'_1$ modes (both around \unit{250}{cm^{-1}} $\widehat{=}$ \unit{31.0}{meV}) and the 2LA mode (\unit{261}{cm^{-1}} $\widehat{=}$ \unit{32.4}{meV}) \cite{NanoLett.16.2363}. (b) Normalized photoluminescence spectra of the monolayer part for different temperatures. The peak that we attribute to the neutral exciton (X$^0$) shows a typical red-shift and an increasing relative amplitude with respect to the other peaks for higher temperatures \cite{APL.105.101901,PhysRevB.90.161302,ScientificRep.6.22414}.}
\label{Raman-PL}
\end{figure*}

\subsection{Energetic upconversion}
The fact that we can pump states significantly lower than the one we probe and nevertheless can measure a clear Kerr rotation signal was explained by the formation of dark trion states in the main manuscript. Nevertheless, we may argue that there might also be a possible energy upconversion of excited states. In principle, such an upconversion has already been observed in WSe$_2$ monolayers, where phonon-mediated anti-Stokes processes enable the scattering from the lower-energy trion state into the higher energy exciton state \cite{NatPhys.12.323}. But as the energy separation between both trion states of \unit{7}{meV} is much larger than the thermal energy at \unit{5}{K}, a much smaller amplitude of the higher energy peak compared to the lower energy peak would be expected in case of such an anti-Stokes process, which is not the case in our experiment, where the Kerr rotation has comparable amplitudes for the cases $E_{\text{pump}}=E_{\text{probe}}$ and $E_{\text{pump}}=E_{\text{probe}}-\unit{7}{meV}$ (see Fig.~2(c) and 2(d) in the main manuscript).

\subsection{Possible doping-induced suppression of a trion-exciton coupling}
Another possible requirement for the long lifetimes observed in our sample may be a doping-induced suppression of a trion-exciton coupling. As long as there exists an effective coupling between trion and free exciton states, a presumably longer lifetime of the trion state might be masked by the one of the exciton. Considering this possibility, we note that the binding energy of the trion state can match the energy of the $A'_1$ optical phonon in WSe$_2$. This enables an anti-Stokes scattering process from the trion into the exciton state. Ref.~\cite{NatPhys.12.323} reports this kind of energy upconversion and states that this scattering process is especially effective if the trion state is resonantly excited, which is exactly the case in our TRKR measurements. However, the reported energy difference between exciton and trion states in Ref.~\cite{NatPhys.12.323} matches perfectly the energy of the $A'_1$ mode (\unit{250}{cm^{-1}} $\widehat{=}$ \unit{31.0}{meV}, see also Raman measurements of our sample in Fig.~\ref{Raman-PL} of this Supplemental Material), whereas the trion binding energy in our case is much larger (\unit{37\pm1}{meV} as seen in Fig.~1(b) of the main manuscript), which can be explained by the high n-doping of our sample \cite{NatNano.8.634, NatNano.9.268}. But a detuning between trion binding and phonon energies most likely renders the phonon-assisted scattering process inefficient \cite{NatPhys.12.323} and, hence, the coupling between trion and exciton states might be suppressed in our case.

\subsection{fs-laser pulses in WSe$_2$ and WS$_2$}
In the main manuscript, we made the notion that a small spectral width of the used laser pulses may be a necessary prerequisite for the observation of long-lived excitonic spin components. In this respect we note that laser pulses in the fs-range have a spectral width in the same order than the observed energy splitting between singlet and triplet states in WSe$_2$. The simultaneous excitation of triplet and singlet states may be one reason why all studies, which used fs-laser pulses, report only ps-lifetimes in WSe$_2$ \cite{PhysRevB.90.161302,NatureComm.6.896,arXiv150704599Y}. A little bit different is the case for WS$_2$: Here, lifetimes in the lower ns-regime are also observed for fs-laser pulses \cite{NatPhys.11.830,arXiv160203568B}, but these values are still much smaller than the tens of ns observed by us or in Ref.~\cite{NanoLett.16.5010}. Furthermore, the splitting between singlet and triplet states in WS$_2$ is reported to be \unit{9}{meV} \cite{NatureComm.7.12715} which is larger than the \unit{7}{meV} observed in WSe$_2$ by us or Ref.~\cite{NatPhys.12.323}.

\subsection{Alternative explanation for the peaks in the PL spectrum}
As dark excitonic states seem to play an important role in the analysis of our data, we want to note that there exist an alternative interpretation of the PL spectrum shown in the main manuscript. There, the peaks at \unit{1.709}{eV} and \unit{1.680}{eV} are attributed to the trion state and a localized state, respectively. Nevertheless, the peaks below the exciton peak at \unit{1.746}{eV} can alternatively be explained as transitions of indirect bright exciton states \cite{PhysRevB.92.125431,PhysRevB.94.075421}. Interestingly, the position of the first two lower-energy peaks exactly match the predictions of reference \cite{PhysRevB.92.125431}. But we note that the positions of the other peaks do not agree to the predictions of reference \cite{PhysRevB.92.125431} and that a newer theoretical study which includes short-range Coulomb exchange interaction in its calculations yields quite different binding energies \cite{PhysRevB.93.121107}. Finally, the assumptions that the PL-peak around \unit{1.709}{eV} is caused by an indirect exciton and not a trion, also contradicts to the fact that we can resolve the trion singlet-triplet fine structure which is a fundamental requirement of our model.

\subsection{Localized excitons and spin signal}
Next, we want to discuss the possibility that the long excitonic spin lifetimes in our sample are caused by localized excitons. Both in our study and the one of Ref.~\cite{NanoLett.16.5010}, which also reports lifetimes of tens of ns, the PL spectrum is dominated by localized states and exhibits an extremely small free exciton peak. TRPL measurements demonstrate that localized states at impurities or defects of TMDC monolayers have a longer lifetime than free excitons \cite{NatPhys.11.477, PhysRevX.6.021024, PhysRevB.90.075413}. But even these longer lifetimes fall short to the observed \unit{150}{ns} in our sample: The longest ones are in the order of one ns \cite{PhysRevB.90.075413, PhysRevB.93.205423} and there is one single study claiming \unit{12}{ns} \cite{PhysRevB.90.155449}. Furthermore, the localized states normally vanish from the PL spectrum in a temperature range between \unit{60-120}{K} \cite{PhysRevB.90.161302, APL.105.101901, ScientificRep.6.22414, Nanoscale.7.10421, PhysRevB.90.075413}. In our case even at \unit{100}{K} the amplitudes of the PL-peaks caused by localized states are comparable with the one of the free exciton (see Fig.~\ref{Raman-PL}(b) of this Supplemental Material). This is in contrast to the temperature dependence of the long-lived excitonic spin component shown in the main manuscript where the long-lived component already vanishes below \unit{50}{K}. Finally, we note that we probe the trion states energy-resolved in our TR Kerr spectroscopy measurements. Hence, there seems to be no \textit{direct} link between localized states and long excitonic spin lifetime. The only indirect link may be that the impurities and defects, which create the localized exciton states, are most likely relevant in the scattering process of the bright singlet trion into the dark state, which is a key requirement of our model explained in the main manuscript.

\subsection{Spot position in magnetic field and temperature measurements}
As mentioned in the main manuscript, the whole cryostat moves a little bit when applying a magnetic field. In temperature dependent measurements also the thermal expansion of the materials of the cryostat change the effective laser spot position on the sample. This is important as PL and Kerr-measurements already demonstrated that a TMDC flake can have spatially-varying optical properties \cite{arXiv160203568B}. To account for this effect, we routinely conducted spatially-resolved measurements of the Kerr- and reflectivity signals. With the help of such maps we repositioned the cryostat to maintain approximately the same laser spot position on the flake for all measurements.

\subsection{Additional data}
\begin{figure*}[tb]
	\includegraphics{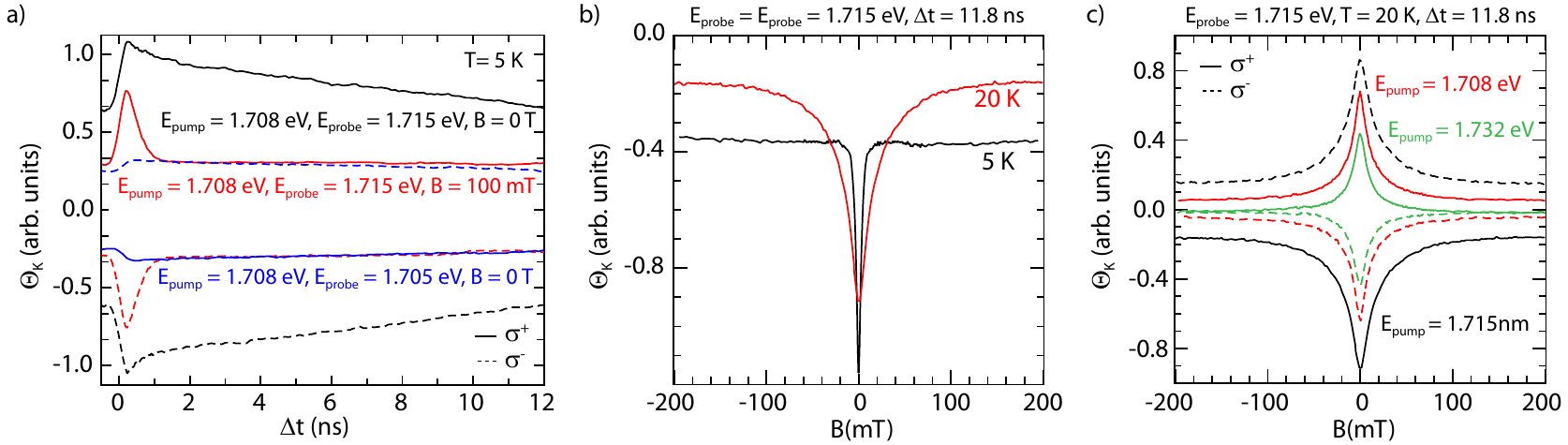}
	\caption{(a) TRKR curves for different laser energies and magnetic fields shown both for $\sigma^+$ (solid lines) and $\sigma^-$ (dashed lines) polarized pump pulses. The change in helicity leads to a clear sign reversal of all three components of the Kerr signal. (b) Resonant spin amplification (RSA, \cite{PhysRevLett.80.4313}) measurements for pump and probe energies in the trion regime at two different temperatures with a delay of $\Delta t=\unit{11.8}{ns}$ between pump and probe pulses. Because individual spin components are either independent of the magnetic field (which leads to the offset in this measurement) or undergo an inhomogeneous spin dephasing, no resonant spin amplification occurs at higher magnetic fields and hence only a single peak can be seen around $B=\unit{0}{T}$. With higher temperatures the full-width-half-maximum of the RSA curve increases and the amplitude of the magnetic field independent contribution decreases. (c) Helicity dependent RSA measurements for several combinations of pump and probe energies.}
	\label{helicity-RSA}
\end{figure*}

\begin{figure*}[tb]
	\includegraphics{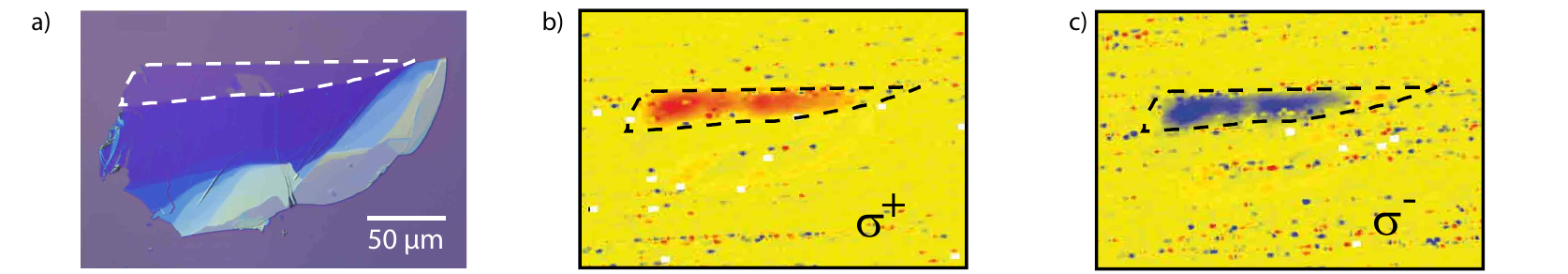}
	\caption{(a) Optical image of the WSe$_2$ flake. The monolayer part is indicated by the dashed line. Only here a Kerr rotation signal can be measured that changes sign between $\sigma^+$ (b) and $\sigma^-$ (c) polarized pump pulses.}
	\label{Kerr-monolayer}
\end{figure*}

\begin{figure*}[tb]
	\includegraphics{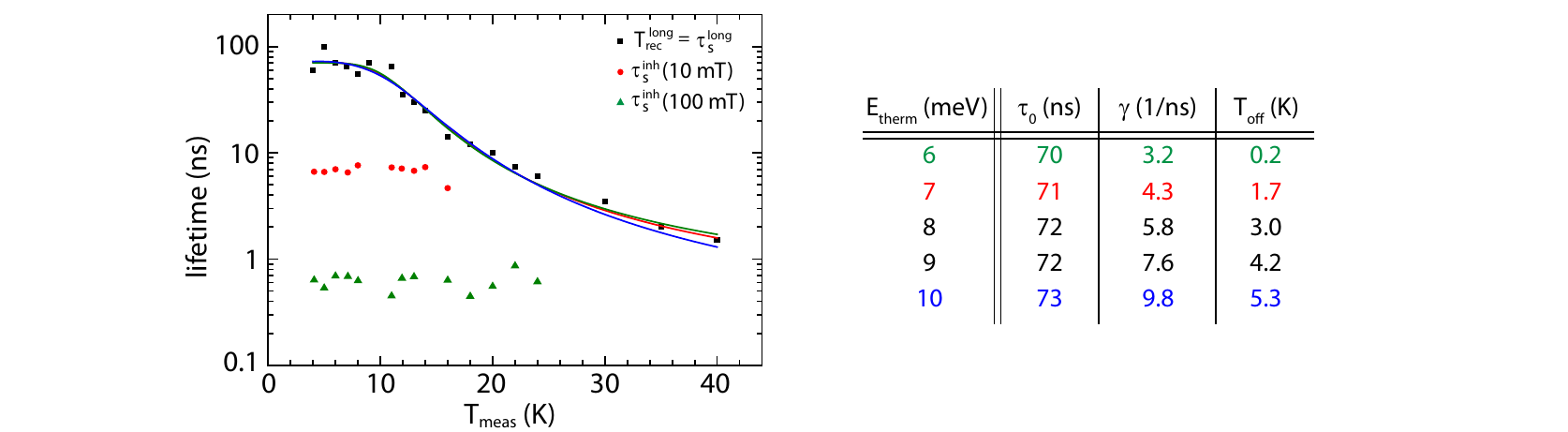}
	\caption{Fit of the long-lived excitonic spin component as a function of temperature with equation 5 of the main manuscript. The thermal activation energy $E_{\text{therm}}$ was set to different values in the range of \unit{6}{meV} to \unit{10}{meV}. The color of the lines correspond to the assigned values in the table. An increase of the activation energy can be compensated by an increase of the thermally activated dephasing rate $\gamma$ and an increase in the assumed difference $T_{\text{off}}$ between the actual sample temperature $T$ and the measured temperature $T_{\text{meas}}$. Our sample was mounted inside a vacuum chamber to a copper block, which is cooled by liquid helium. The temperature sensor is situated inside this copper block. Laser-induced heating and a finite thermal conductance between sample and copper will lead to a small temperature gradient between the sample and the copper block.}
	\label{thermal-activation}
\end{figure*}

\begin{figure*}[tb]
	\includegraphics{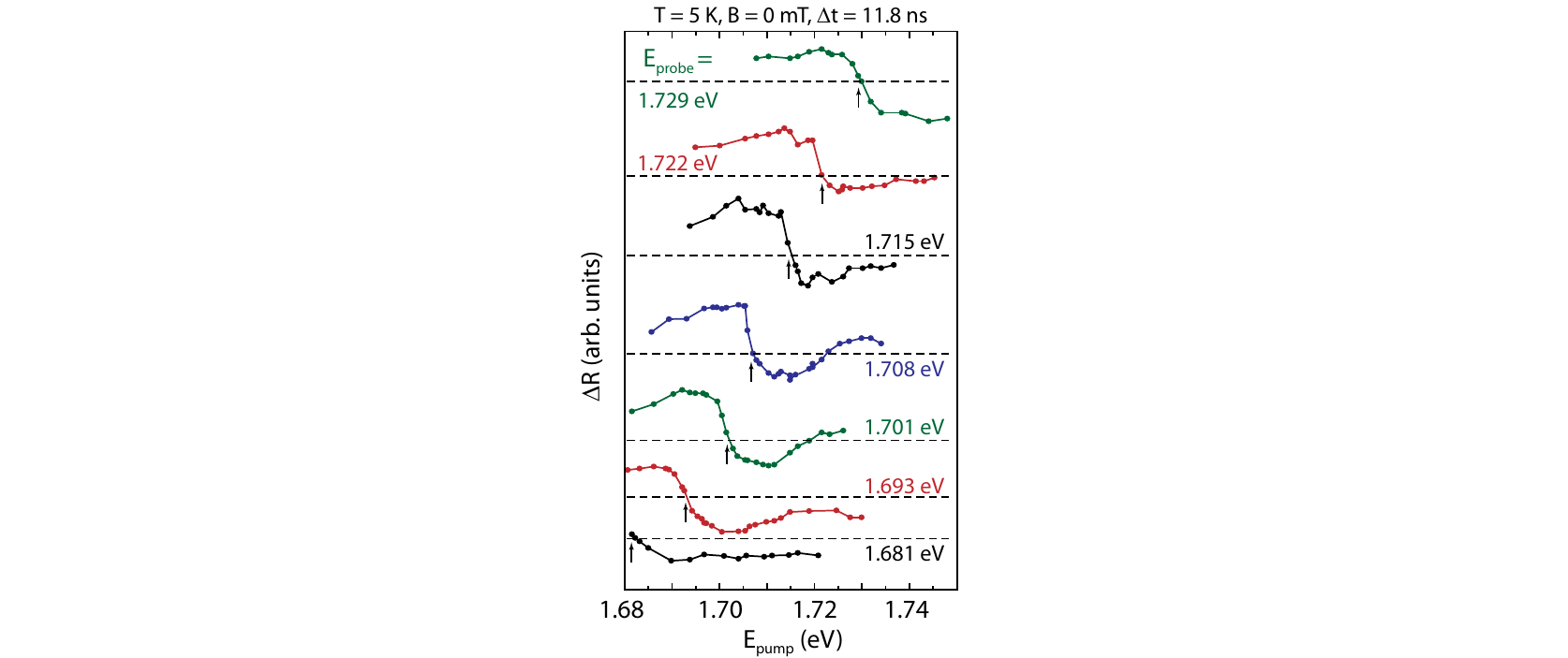}
	\caption{Change in pump-induced reflectivity as a function of pump energy for different probe energies. The data was acquired simultaneously to the Kerr rotation data of figure 2(d) of the main manuscript. The dashed lines are guides to the eye and go through the point where pump and probe energy are equal (also indicated by black arrows). As in the case of the Kerr rotation data the resonance in the reflectivity signal shifts with probe energy. Furthermore, the reflectivity exhibits a similar pronounced edge at $E_{\text{pump}}=E_{\text{probe}}$ like the ellipticity signal in figure 2(b) of the main manuscript.}
	\label{reflectivity}
\end{figure*}

\begin{figure*}[p]
	\includegraphics{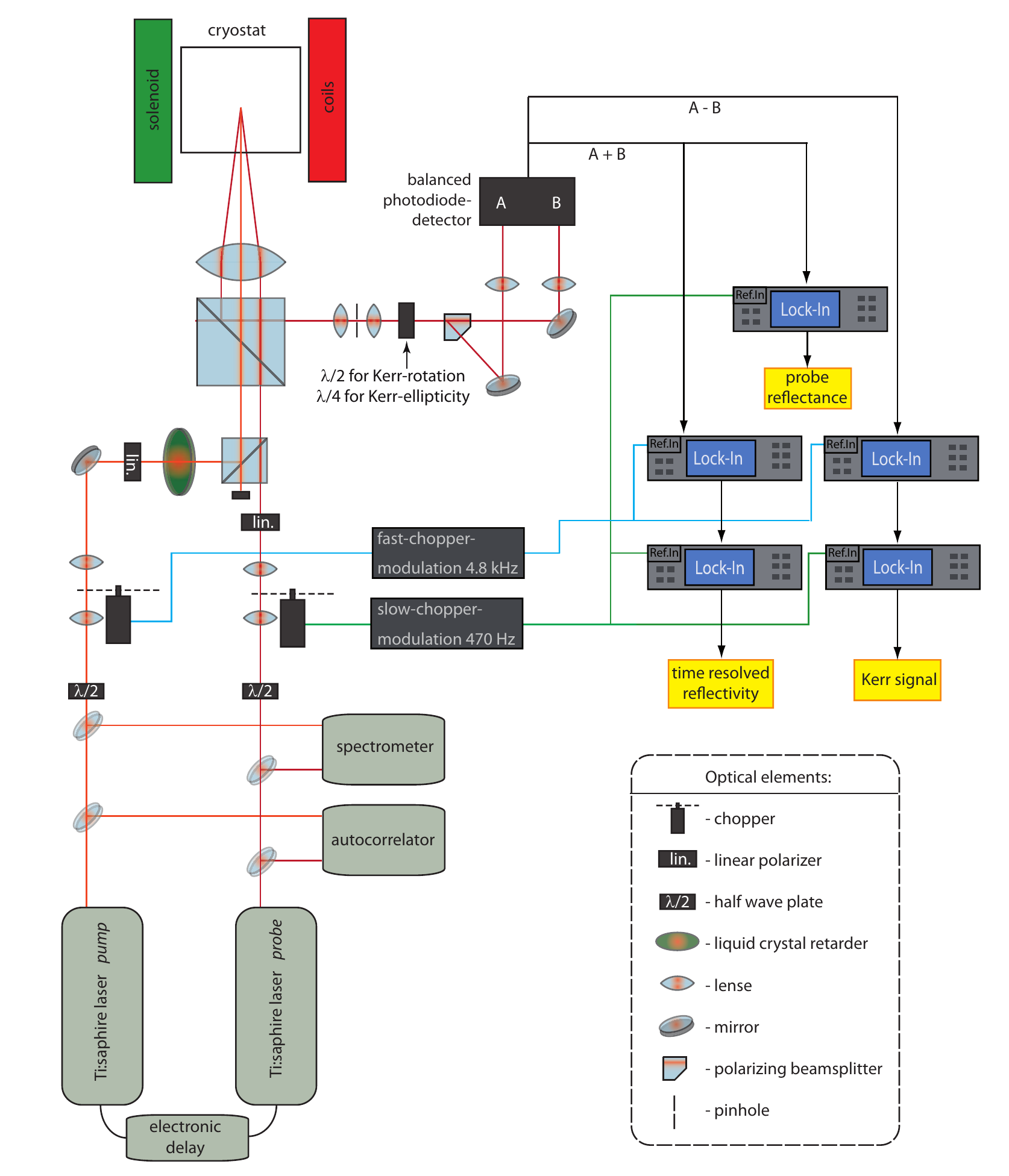}
	\caption{Scheme of the experimental setup. Pump and probe laser pulses stem from two independently tunable Ti-sapphire ps lasers, which are synchronized and electronically delayed with a temporal jitter of less than 1 ps. The Kerr rotation signal is detected by a polarization bridge with balanced photodiodes \cite{PhysRevB.56.7574}.}
	\label{setup}
\end{figure*}

For the rest of this Supplemental Material we show Kerr rotation signals for different helicities and RSA-measurements in Fig.~\ref{helicity-RSA}, a map showing the spatial distribution of the Kerr rotation signal in Fig.~\ref{Kerr-monolayer}, additional fittings of the thermal activation energy in Fig.~\ref{thermal-activation}, reflectivity data for different pump/probe energies in Fig.~\ref{reflectivity}, a schematic drawing of our setup in Fig.~\ref{setup}, and an overview of the results of previous Faraday/Kerr-studies on TMDCs in the table at the end of this Supplemental Material.

\clearpage
\newpage

\begin{sidewaystable}[p]
\vspace{8.5cm}
	\resizebox{\textwidth}{!}{
	    \begin{tabular}{|l|r|l|r|l|c|r|}
    \hline
          &       & longest lifetime and form of &       &       &       & \multicolumn{1}{l|}{transfer of spins}  \\
    reference & \multicolumn{1}{l|}{material} & Kerr/Faraday signal & \multicolumn{1}{l|}{magnetic field dependence} & reflectivity/transmittivity & pulse width & \multicolumn{1}{l|}{to resident carriers?}  \\
    \hline
    Zhu et al. & \multicolumn{1}{l|}{WSe2} & exciton: 6 ps @ 4 K & \multicolumn{1}{l|}{n.a.} & 10x longer than Kerr signal & 120 fs & \multicolumn{1}{l|}{no evidence }  \\
    \cite{PhysRevB.90.161302} &       & mono-exponential  &       &       &       &   \\
    \hline
    Hsu et al.  & \multicolumn{1}{l|}{WSe2} & trions: 700 ps @ 10 K & \multicolumn{1}{l|}{n.a.} & similar to Kerr signal & 150 fs & \multicolumn{1}{l|}{yes}  \\
    \cite{NatureComm.6.896} & \multicolumn{1}{l|}{p-doped} & excitons 10 ps @ 10 K &       & three exponential components &       &  \\
          &       & mono-exponential &       &       &       &   \\
    \hline
    Song et al.  & \multicolumn{1}{l|}{WSe2} & exciton: 80 ns @ 10 K & \multicolumn{1}{l|}{no magnetic field dependence} & n.a.  & 3 ps  & \multicolumn{1}{l|}{yes}  \\
    \cite{NanoLett.16.5010} & \multicolumn{1}{l|}{p-doped} & three exponential decays &       &       &       &   \\
    \hline
    Volmer et al.  & \multicolumn{1}{l|}{WSe2} & trions: 150 ns @ 5 K & \multicolumn{1}{l|}{dependent on component: } & equal to two of the Kerr components & 3 -100 ps & \multicolumn{1}{l|}{no }  \\
    (this paper) & \multicolumn{1}{l|}{n-doped} & four different components & \multicolumn{1}{l|}{no field dependence, inhomogeneous } & bi-exponential &       &  \\
          &       &       & \multicolumn{1}{l|}{dephasing or oscillatory signal} &       &       &   \\
    \hline
    Yan et al. & \multicolumn{1}{l|}{WSe2} & exciton: 4 ps @ 10 K & \multicolumn{1}{l|}{n.a.} & exciton: longer  than Kerr signal & 150 fs &   \\
    \cite{arXiv150704599Y} &       & trion: 123 ps @ 10 K  &       & trion: shorter than Kerr signal &       &  \\
          &       & mono-exponential for excitons &       & bi-exponential &       &  \\
          &       & bi-exponential for trions &       &       &       &   \\
    \hline
    Yan et al.  & \multicolumn{1}{l|}{WSe2} & 2 ps to 2 ns @  70 K & \multicolumn{1}{l|}{n.a.} & longer or shorter than Kerr signal & n.a.  & \multicolumn{1}{l|}{yes}  \\
    \cite{arXiv161201336Y} &       & mono- or bi-exponential  &       & depending on &       &  \\
          &       & depending on pump/probe energy &       & pump/probe energy &       &   \\
    \hline
    Plechinger et al.  & \multicolumn{1}{l|}{WS2} & B exciton: 800 ps @ 4.5 K & \multicolumn{1}{l|}{n.a.} & shorter than Kerr signal & 180 fs &   \\
    \cite{NatureComm.7.12715} &       & A exciton: ps-range @ 4.5 K &       & bi-exponential &       &  \\
          &       & bi-molecular fit plus &       &       &       &  \\
          &       & additional short component &       &       &       &   \\
    \hline
    Bushong et al. & \multicolumn{1}{l|}{WS2} & exciton: 5.4 ns @ 6 K & \multicolumn{1}{l|}{signal robust against external} & n.a.  & 150 fs & \multicolumn{1}{l|}{yes}  \\
    \cite{arXiv160203568B} & \multicolumn{1}{l|}{ n-doped} & bi-exponential fit &   \multicolumn{1}{l|}{magnetic fields}    &       &       &   \\
    \hline
    Yang et al. & \multicolumn{1}{l|}{WS2, MoS2} & 3 ns @ 5 K & \multicolumn{1}{l|}{decreasing lifetime at higher field} & ns-range & 250 fs & \multicolumn{1}{l|}{yes}  \\
    \cite{NatPhys.11.830} & \multicolumn{1}{l|}{n-doped} & exponential for B = 0 T & \multicolumn{1}{l|}{oscillatory signal at high fields} &       &       &   \\
    \hline
    Yang et al. & \multicolumn{1}{l|}{MoS2} & 3 ns @ 5 K & \multicolumn{1}{l|}{inhomogeneous spin dephasing} & n.a.  & 250 fs & \multicolumn{1}{l|}{yes}  \\
    \cite{NanoLett.15.8250} & \multicolumn{1}{l|}{n-doped} & fast decay + long lived  &       &       &       &  \\
          &       & spin precession &       &       &       &   \\
    \hline
    Dal Conte et al.  & \multicolumn{1}{l|}{MoS2} & exciton: 5 ps @ 77 K & \multicolumn{1}{l|}{n.a.} & equal to Faraday signal & 70 fs &   \\
    \cite{PhysRevB.92.235425} &       & bi-exponential fit &       & bi-exponential &       &   \\
    \hline
    \end{tabular}}
\end{sidewaystable}%

\end{document}